\documentclass[aps,prb,reprint,superscriptaddress,floatfix]{revtex4-2}

\usepackage[utf8]{inputenc}
\usepackage{amsmath}
\usepackage{amsfonts}
\usepackage{amssymb}
\usepackage{graphicx}
\usepackage{physics}
\usepackage{siunitx}
\usepackage{float}
\usepackage{subcaption}
\usepackage{booktabs}
\usepackage{multirow}

\usepackage{adjustbox}

\usepackage{xr-hyper}  
\usepackage{hyperref} 

\usepackage{pdfpages}
\usepackage{pgffor}

\makeatletter
\AtBeginDocument{\let\LS@rot\@undefined}
\makeatother

\begin{document}

\title{Origin of Flat Bands and Role of Electron Correlation in Lutetium Hydrides}

\author{Anmol Lamichhane}

\affiliation{Department of Physics, University of Illinois Chicago, Chicago, IL}

\author{Adam Denchfield}
\affiliation{Oden Institute for Computational Engineering and Sciences, The University of Texas at Austin, Austin, Texas 78712, USA}

\author{Hyeondeok Shin}
\affiliation{Computational Science Division, Argonne National Laboratory, Lemont, Illinois 60439, USA}

\author{P. Ganesh}
\affiliation{Center for Nanophase Materials Sciences, Oak Ridge National Laboratory, Oak Ridge, TN}

\author{Russell J. Hemley}
\affiliation{Department of Physics, University of Illinois Chicago, Chicago, IL}
\affiliation{Department of Chemistry, University of Illinois Chicago, Chicago, IL}
\affiliation{Department of Earth and Environmental Sciences, University of Illinois Chicago, Chicago, IL}
\affiliation{Materials Science Division, Argonne National Laboratory, Lemont, Illinois 60439, USA}

\author{Hyowon Park}
\affiliation{Department of Physics, University of Illinois Chicago, Chicago, IL}
\affiliation{Materials Science Division, Argonne National Laboratory, Lemont, Illinois 60439, USA}

\date{July 20, 2026}

\begin{abstract}
Lutetium hydrides (LuH$_x$, $1.75 \leq x \leq 3$) form a diverse series of
phases, several of which superconduct under pressure. Characterizing their
electronic properties has remained challenging owing to a high propensity for
hydrogen defect formation, and recent angle-resolved photoemission (ARPES)
measurements reveal puzzling flat-band regions that position these materials as
candidates where superconductivity and flat-band physics may intersect. Here, by combining density functional theory, dynamical mean-field theory, and the constrained random-phase approximation, we uncover the microscopic origin and correlation nature of these flat bands. Across all compositions, the screened on-site Coulomb interaction is larger for H-s states than for Lu-$d$ states due to compact hydrogen orbitals. Nevertheless, these systems remain weakly correlated metals: the nearly filled H-$s$ shell admits
little charge fluctuation, so its large interaction acts as a static level shift
rather than a source of correlation. Although hydrogen primarily occupies tetrahedral sites at $x=2$, we discover that anti-site defects—where hydrogens occupy slightly unfavorable octahedral sites—generate both the ARPES flat-band features and the low-energy optical absorption peak, attesting to the usual defective nature of such materials in experimental samples. We further find that correlation strength is governed primarily by hydrogen orbital filling at these sites rather than the interaction magnitude itself. Consequently, we identify hydrogen orbital filling as the fundamental organizing principle dictating correlation and low-energy flat-band physics in lutetium hydrides.  
	
\end{abstract}

\maketitle

\section{Introduction}

Many metal hydrides are well described within conventional band theory, as the metal-derived electrons are mostly itinerant and they effectively screen the hydrogen states.  The coupled electronic, structural and phonon properties of stoichiometric and non-stoichiometric lanthanide hydrides have been a puzzling topic of study for over half century. Motivated by the argument of Matthias \cite{Matthias1957}, Merriam and Schreiber~\cite{MerriamSchreiber1963} searched for  superconductivity in lanthanum hydrides to below 1 K. Two decades later, a sub-kelvin $T_c$ for LaH$_2$ was estimated based on Bardeen--Cooper--Schrieffer (BCS) theory~\cite{GuptaBurger1980}. This superconductivity was ultimately confirmed ~\cite{Kai1989} in slightly superstoichiometric
LaH$_{2+x}$. A growing body of later work suggested that certain rare-earth hydrides exhibit behavior beyond the standard band-theory description, including
electron--electron correlation effects and correlation-driven metal--insulator transitions.\cite{huiberts_yttrium_1996,ng_electronic_1997,shinar_q-factor_1988,Libowitz1972} This beyond-band-theory behavior originates in hydrogen lattice itself: a periodic array of such correlation-stabilized ions
is a genuine many-body problem exhibiting metallic, intermediate, and
Mott-insulating regimes~\cite{ZgidChan2011}, and correlation-based treatments have
accordingly cast LaH$_3$/YH$_3$ as correlation-driven
insulators~\cite{Ng1999,Eder1997}, with hydrogen ordering proposed as an
additional contributing factor~\cite{Vajda2004,Kerscher2012}.

Lutetium hydride is a particularly clean system in which to test these ideas.
Its filled $4f$ shell removes the $f$-electron correlation that complicates the
lighter rare earths, isolating the hydrogen and Lu-$5d$ states as the only
candidates for correlated behavior; and its accessible dihydride--trihydride
range spans hydrogen fillings from the tetrahedral-only dihydride to the
octahedrally-filled trihydride, allowing the hydrogen occupation to be tuned
systematically.  Weaver et al.~\cite{Weaver1979optical} measured optical reflectance spectra that established the dominance of Drude absorption  and yielded the carrier plasma frequency, while synchrotron photoemission resolved the Lu-$d$ conduction states near the Fermi level and a H-$s$ band several eV below it, providing evidence for charge transfer from the lutetium conduction ($5d/6s$) manifold to the hydrogen site~\cite{Weaver1979pes}. The same measurements locate the filled Lu $4f$ levels as sharp features near $-8$~eV;
being a closed, chemically inert shell, the $4f$ states do not participate in
the bonding with hydrogen but distinguish lutetium from lanthanum, whose empty $4f$
shell underlies the larger lanthanum lattice constant.
The electronic structure of the lutetium dihydride has been
characterized experimentally: angle-resolved photoemission on Lu-H-N have been interpreted as indicating a flat band $\sim0.2$~eV below the Fermi
level~\cite{liang_observation_nodate}, which has been suggested to reflect
strong correlation; whether the near-$E_F$ flat band is genuinely
correlation-driven, however, remains an open question.

Further, theoretical studies of Sufyan and Larson ~\cite{Sufyan2023} report that pristine LuH$_3$, in its trigonal structure, hosts nontrivial topological features
and van Hove singularities near the Fermi level, and Denchfield et al.~\cite{Denchfield2024} show that specific substitutions of nitrogen in the hydrogen sublattice generates flat bands and
sharply peaked densities of states.

We show that electronic correlation in the rare-earth hydrides is governed
by hydrogen occupation rather than interaction strength --- the central
result of this work, and one that inverts the usual expectation. Throughout,
we use \emph{correlation strength} to mean the magnitude of many-body
renormalization of the one-particle properties, quantified by the mass
enhancement $m^*/m = 1/Z$ and by the local spin susceptibility $\chi_S$ and
instantaneous moment $\langle m^2 \rangle$; we reserve \emph{interaction strength} for the magnitude of the screened on-site $U$ and $J$ obtained from cRPA. The two need not track one another, and in these materials they
do not: hydrogen carries large cRPA interaction yet remains the weakly renormalized site. Moreover, two hydrogen sublattices are electronically inequivalent, with octahedral hydrogen the more strongly hybridizing of the two, and we find that specific arrangements of tetrahedral vacancies and octahedral occupancies give rise to narrow bands and van Hove singularities near $E_F$.

The remainder of this paper is organized as follows. Section II presents
the results: the electronic inequivalence of the tetrahedral and
octahedral hydrogen sublattices, the mass enhancements and local moments
obtained from DMFT, the DFT and DMFT densities of states, the effects of
octahedral occupancy and hydrogen vacancies, and comparison with
experiment. Section III discusses the origin of the hydrogen filling and
its relation to the observed renormalization, and compares our results
with spectroscopic measurements on lutetium and other lanthanide
hydrides. The paper concludes with a summary of how hydrogen occupation
controls the strength of correlation effects in rare-earth hydrides.

\section{Results}

\subsection{Inequivalent Tetrahedral and Octahedral Hydrogen Sublattices}
\label{sec:results-onsite}

All LnH$_x$ structures considered here (Ln = Lu, La) are built on the cubic
fluorite (CaF$_2$-type) framework, space group $Fm\bar{3}m$. The metal atoms
form a face-centered cubic (FCC) sublattice on the $4a$ sites, and hydrogen occupies the
tetrahedral $8c$ sites, each
coordinated by four metal atoms. This accounts for two hydrogen per formula
unit, giving the dihydride LnH$_2$. The remaining octahedral interstitial, the
$4b$ site and coordinated by six
metal atoms, is vacant in the ideal fluorite structure; progressive filling of
this site carries the composition from $x = 2$ toward the trihydride LnH$_3$, in
which both hydrogen sublattices are fully occupied.  We begin by asking whether the two hydrogen sublattices in LnH$_3$ are electronically equivalent. Figure~1 shows the model structure for LuH$_3$, with tetrahedral hydrogen
(H$_{\mathrm{tet}}$) in pink and octahedral hydrogen (H$_{\mathrm{oct}}$) in
blue. 

\begin{figure}[t]
	\centering
	\includegraphics[width=0.8\linewidth]{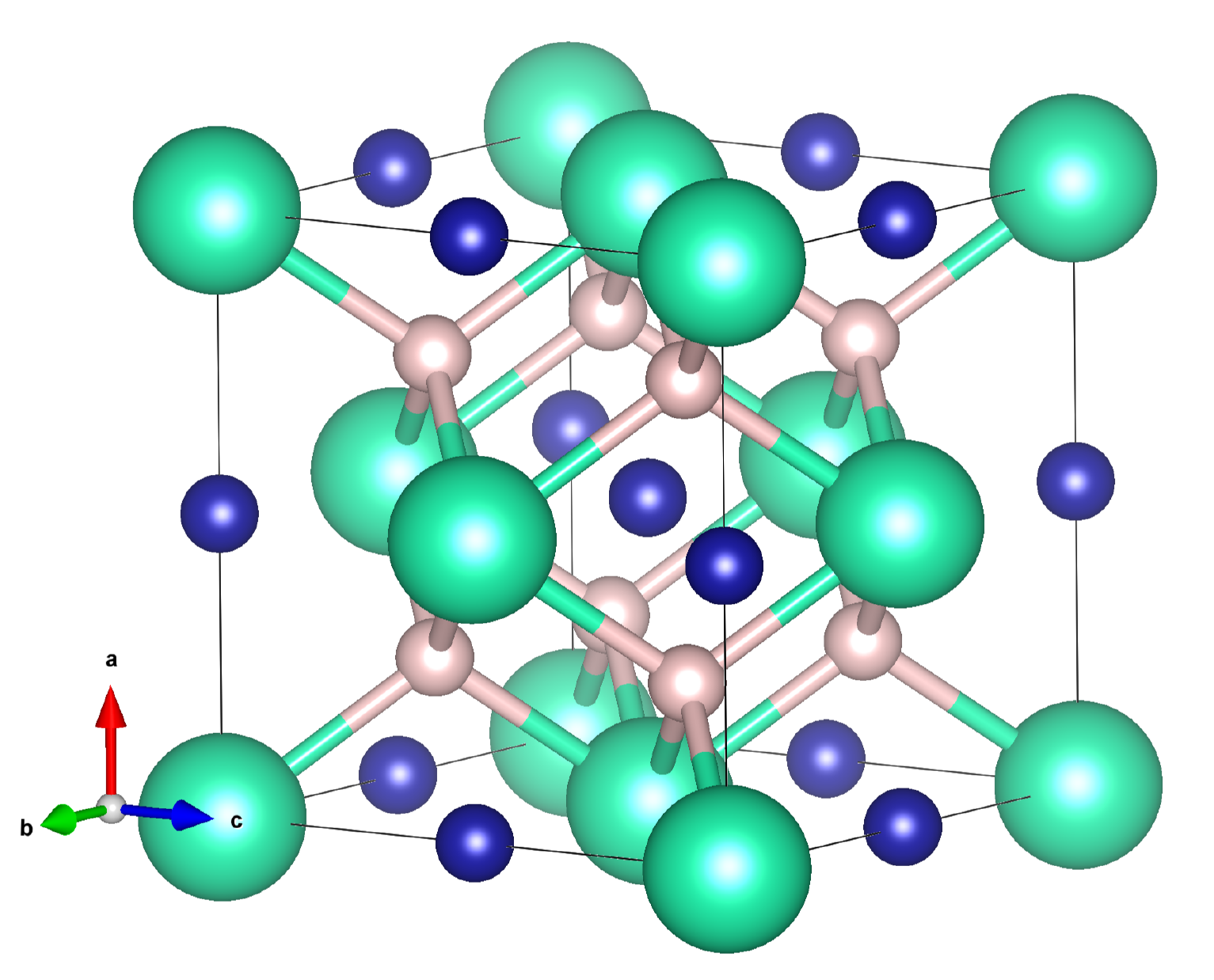}
	\caption{Idealized fluorite structure of LuH$_3$. Tetrahedral hydrogens (H$_{tet}$) are represented by pink atoms, octahedral hydrogens (H$_{oct}$) are represented by blue atoms and Lutetium atoms are represented by turquoise atoms.}
	\label{fig:band_dos}
\end{figure}

To characterize the two hydrogen sites electronically, we extracted their on-site
energies, for LnH$_3$, from the diagonal elements of the Wannier Hamiltonian (Table 1). The tetrahedral and octahedral hydrogen sublattices are found to be electronically
inequivalent, differing substantially in their on-site energies
(Table~\ref{tab:onsite}). For LuH$_3$, the tetrahedral H-$s$
level lies $0.97$~eV below the octahedral H-$s$ level, with both situated well
below the Fermi level. The Lu-$d$ on-site energies, by contrast, lie above
$E_F$, split by the crystal field into two sets at $\epsilon-E_F \approx +2.1$
and $+3.8$~eV, well separated from the hydrogen manifold. The tetrahedral
hydrogen, fully coordinated and filled already in the dihydride, is the more
strongly bound and more deeply lying species; the octahedral hydrogen---the
``extra'' hydrogen added on going from the di- to the trihydride---is more
weakly bound and sits closer to $E_F$.

The ordering is reversed in LaH$_3$, where the octahedral level lies $0.38$~eV
\emph{below} the tetrahedral one. The two compounds differ substantially in
lattice constant ($5.01$ versus $5.65$~\AA), and the octahedral interstitial,
being the larger and less tightly confined site, is expected to be the more
sensitive of the two to this difference; the reversal is consistent with the
site-resolved occupations, which show the octahedral hydrogen to be the more
filled species in LaH$_3$ and the less filled one in LuH$_3$. We do not attempt
to separate the geometric effect from possible chemical differences between the
two rare earths.

\begin{table}[h]
	\centering
	\caption{On-site ($R=0$ diagonal) Wannier energies of the tetrahedral and
		octahedral hydrogen $s$ states relative to the Fermi level, from the MLWF
		Hamiltonian, for LuH$_3$ ($E_F = 4.845$~eV) and LaH$_3$
		($E_F = 7.828$~eV).}
	\label{tab:onsite}
	\begin{tabular}{l c c}
		\hline\hline
		Orbital & \multicolumn{2}{c}{$\epsilon - E_F$ (eV)} \\
		& LuH$_3$ & LaH$_3$ \\
		\hline
		H$_{\mathrm{tet}}$-$s$ & $-3.841$ & $-2.751$ \\
		H$_{\mathrm{oct}}$-$s$ & $-2.868$ & $-3.129$ \\
		\hline\hline
	\end{tabular}
\end{table}

\subsection{Cubic trihydrides: interactions do not open the gap}

\begin{figure}[t]
	\centering
	\includegraphics[width=1.\linewidth]{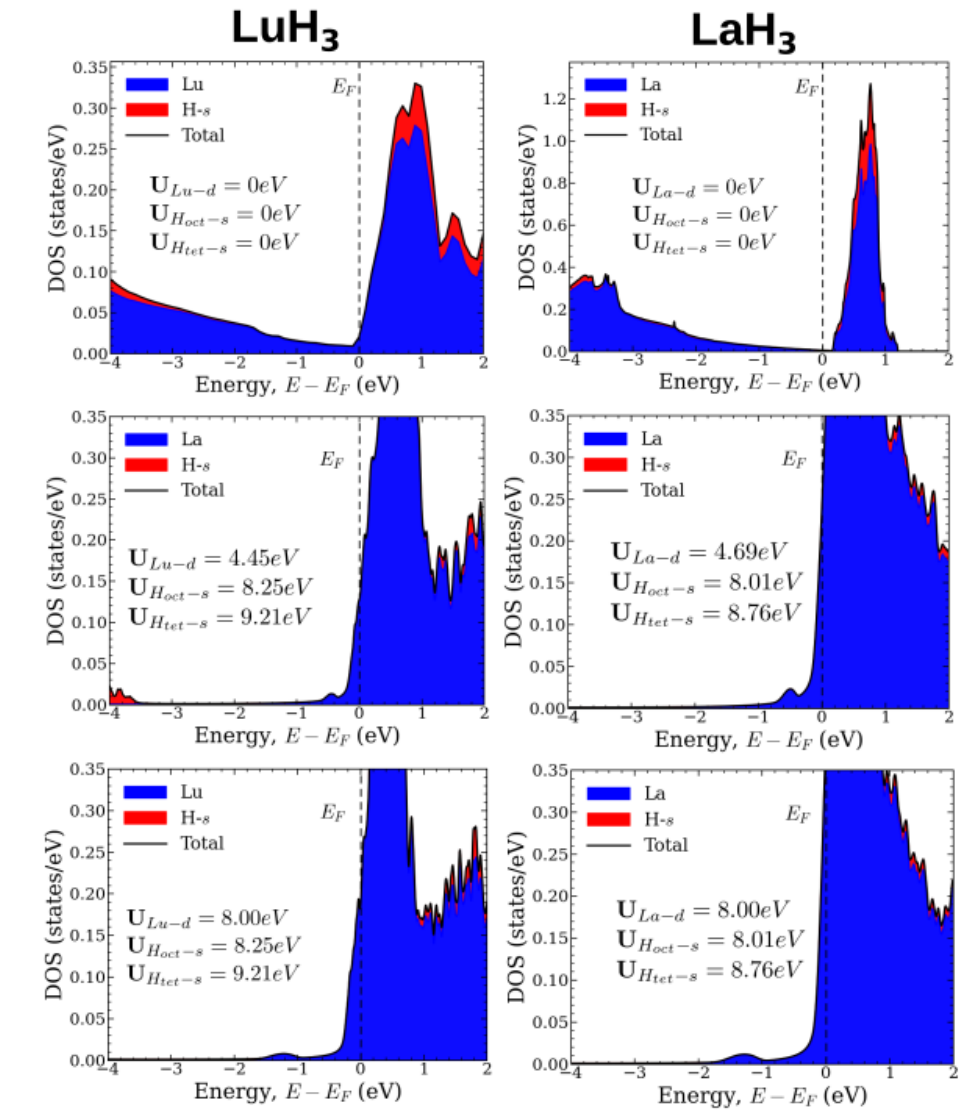}
\caption{\textbf{Hubbard interactions alone do not gap the cubic trihydrides.}
	Orbital-resolved DOS of LuH$_3$ (left) and LaH$_3$ (right) at the Wannier	level ($U=0$, top), from DMFT at the cRPA couplings (middle), and from DMFT with $U_{d}=8$~eV at fixed hydrogen interactions (bottom).}
	\label{fig:dos}
\end{figure}

LaH$_3$ and YH$_3$ are  semiconductors~\cite{huiberts_yttrium_1996}, whereas standard band
theory predict these trihydrides to be metallic; hydrogen ordering~\cite{Vajda2004,Kerscher2012} and
electronic correlation ~\cite{Ng1999,Eder1997} have both been proposed as the missing ingredient to open the band gap.
Figure~\ref{fig:dos} tests the correlation scenario directly.

At the Wannier level ($U=0$, top row) LuH$_3$ and LaH$_3$ are low-density-of
-states metals: $E_F$ lies in a shallow minimum with
$N(E_F)=0.02$ and $0.002$~states/eV respectively, produced by residual
overlap between the H-derived valence weight and the RE-$d$ conduction
manifold, and part of the H-$s$ weight remains unoccupied.

Applying correlation at the cRPA couplings (middle row) rearranges the spectrum substantially. The static, orbital-dependent shift of $\mathrm{Re}\,\Sigma(0)-V^{\rm dc}$ drives the occupied H-derived weight
below $-4$ eV and removes the unoccupied H-$s$ admixture,
leaving a wide region of negligible spectral weight beneath $E_F$. Both
compounds nonetheless remain metallic: the $E_F$ is just above
the $d$ conduction manifold with the total density of states  $N(E_F)=0.13$ and $0.27$~states/eV and a
finite $d$ occupation $n_d=0.68$ per rare earth atom. Raising $U_d$ to $8$~eV,
nearly twice the cRPA value (bottom row), leaves this unchanged, only shifting the $d-$ states below $E_F$ by a little amount due to the remaining unoccupied H orbitals.

The occupations identify this orbital dependent correlation effect. Hydrogen is a nearly filled shell
($n_{\mathrm{H}}=1.8$--$1.9$) and the $d$ manifold is nearly empty
($n_d=0.5$--$0.7$ of ten spin-orbitals); the insulating state these
compounds are expected to reach is therefore the ionic RE$^{3+}$(H$^-$)$_3$ band insulator, requiring $n_{\mathrm{H}}\to2$ and
$n_d\to0$. The two $d$ sets are themselves well separated: in LuH$_3$ the $e_g$ doublet lies
at $\epsilon - E_F = +2.08$~eV and the $t_{2g}$ triplet at $+3.81$~eV, and in
LaH$_3$ at $+1.75$ and $+2.75$~eV respectively. The $e_g$ set therefore lies below $t_{2g}$ in both compounds---the inverse of the familiar octahedral
ordering, and the expected result for the eight-fold cubic coordination of the
metal by tetrahedral hydrogen in the fluorite structure.

Hubbard interactions push the system toward the insulator limit but cannot
complete the charge transfer, because the residual $d$ occupation is sustained by
the strong s--$d$ hybridization. Doubling $U_d$ accordingly changes nothing. It however is possible to open the gap through distorting the hydrogen sublattice, consistent with
DFT$+U$ and quantum Monte Carlo results in which hydrogen distortions open
a small gap that interactions subsequently enlarge~\cite{denchfield_correlation_2025}---which motivates the vacancy- and
occupancy-resolved analysis of the following sections.

\subsection{Correlation and Vacancy Effects}
\label{sec:results-correlation}

We now consider the electronic structure of LuH$_x$ at five compositions, $x = 1.75$, $2$, $2.25$, $2.875$, and $3$. The compositions studied here therefore span partial vacancy of the tetrahedral sublattice below $x = 2$ and partial filling of the octahedral sublattice above
it. 
Figure~3 shows the DFT-relaxed structures for the
compositions considered, together with their band structures and densities of
states; the sub-stoichiometric compounds, in which tetrahedral sites are
partially vacant, develop flat bands immediately below the Fermi level. The
DFT-optimized lattice parameters and the cRPA interaction parameters for all
compositions are given in Sections~S2.1 and~S2.2, respectively.

\begin{figure*}[t]
	\centering
	\includegraphics[width=.65\linewidth]{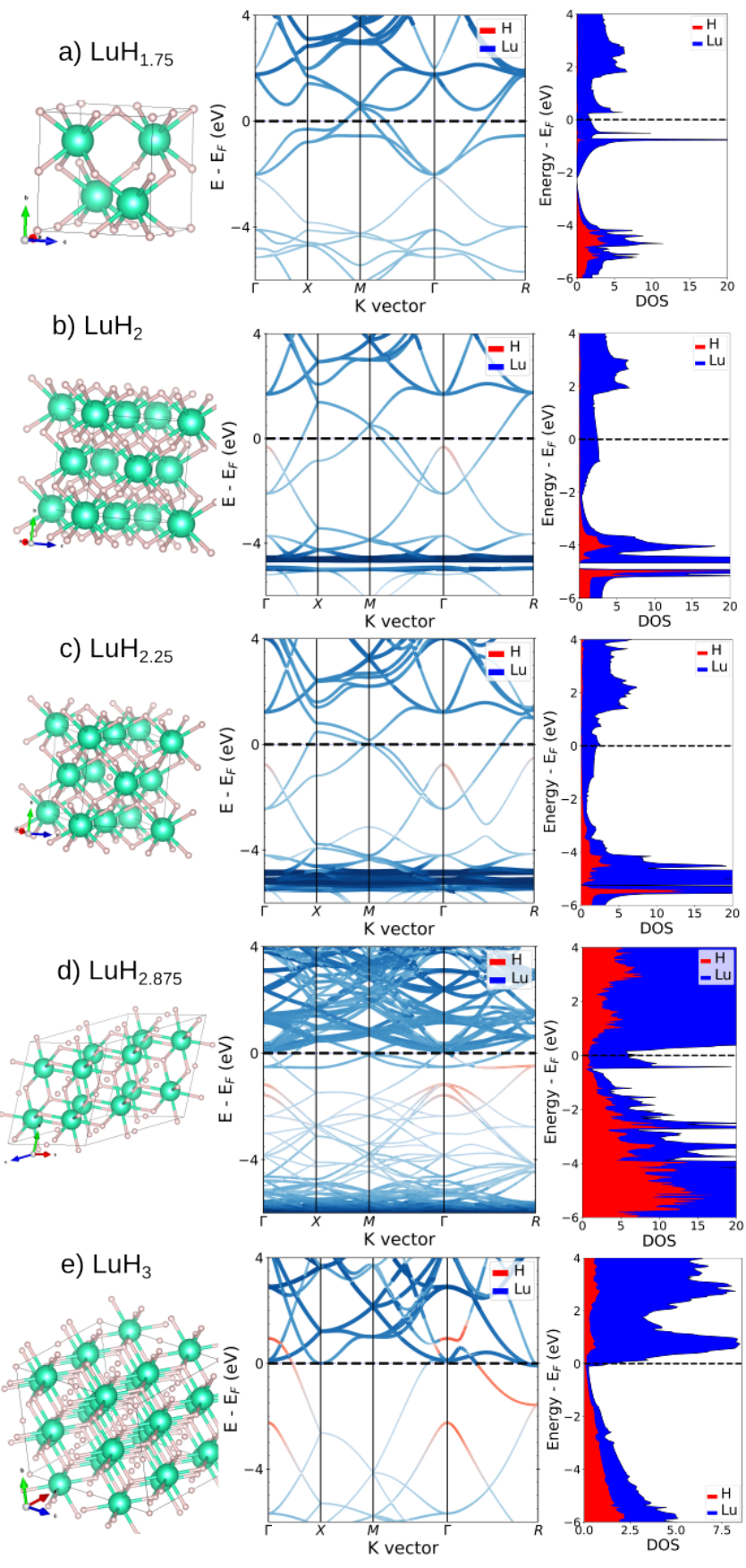}
	\caption{Lu-H structures with different hydrogen occupation of H$_{oct}$ and H$_{tet}$ studied and their associated DFT calculated electronic properties. For 4 Lu atoms: a) LuH$_{1.75}$ has 7 H$_{tet}$. b) LuH$_{2}$ has 8 H$_{tet}$. c) LuH$_{2.25}$ has 8 H$_{tet}$ and 1 H$_{oct}$. d) LuH$_{2.875}$ has 8 H$_{tet}$ and 3 H$_{oct}$. e) LuH$_{3}$ has 8 H$_{tet}$ and 4 H$_{oct}$.}
	\label{fig:band_dos}
\end{figure*}

Across the LuH$_x$ series the hydrogen $s$ shell is close to being fully filled at both
sites ($n = 1.72$--$1.85$), consistent with an H$^-$-like anion, but it is not a
closed shell: the deficit from $n=2$ is recovered almost exactly as excess $d$
occupation (Table 18, $r=0.99$), establishing that the
two sublattices are hybridized rather than ionically decoupled. The hydrogen
sublattices are themselves resolvably inequivalent. At every composition containing octahedral hydrogen, $n_{\mathrm{oct}} = 1.72$--$1.76$ and
$n_{\mathrm{tet}} = 1.80$--$1.85$, which follows directly from the on-site
energies of Sec.~\ref{sec:results-onsite}: the shallower octahedral level is
less completely occupied than the deeper tetrahedral one.

Despite the predominantly closed-shell character, residual charge fluctuations are substantial: $P_{\mathrm{double}} = 0.73$--$0.85$, which corresponds to non-closed-shell configuration $15$--$27\%$ of the time. The associated instantaneous
moment reaches $\langle m^2\rangle = 0.26$ for the octahedral sites, versus $0.145$--$0.19$ on tetrahedral ones. For a single orbital these are not
independent diagnostics: $\langle m^2\rangle = 2-n-2P_{\mathrm{empty}}$
identically, and with $P_{\mathrm{empty}} < 0.012$ throughout, the moment is a
direct readout of the charge deviation from the ionic limit. Local charge and
spin fluctuations therefore measure the hybridization strength site by site,
and are largest where the on-site level lies closest to $E_F$.

The hydrogen mass enhancement increases systematically with octahedral
occupation. At the tetrahedral site it rises monotonically from
$m^*/m = 1.20$ in LuH$_{1.75}$ through $1.22$ in LuH$_2$ and $1.29$ in
LuH$_{2.875}$ to $1.34$ in LuH$_3$, and the octahedral site reaches $1.36$ in
LuH$_3$. These values of mass enhancement are comparable to the kagome metal ScV$_6$Sn$_6$
($m^*/m \approx 1.3$)~\cite{Yu2024ScV6Sn6} and well below the value for moderately correlated
oxides such as SrVO$_3$ ($m^*/m \approx 2$)~\cite{Yoshida2005}.

The Lu-$d$ manifold is similarly weakly correlated but follows a different
trend. Its occupation decreases monotonically with hydrogen content, from
$n_d = 1.59$ in LuH$_{1.75}$ to $0.68$ in LuH$_3$, as hydrogen withdraws charge
from the $d$ band. The $d$ mass enhancement, however, is non-monotonic: it peaks
near the dihydride ($m^*/m \approx 1.28$--$1.32$ for LuH$_2$) and decreases
toward the trihydride ($1.16$--$1.18$ in LuH$_3$). The two crystal-field-split
$d$-sets track each other closely at all compositions.

The local susceptibilities distinguish the two channels more sharply than the static moments alone. The instantaneous moment measures the magnitude of the local moment at a given instant; the local spin susceptibility $\chi_S$ additionally weights its persistence in imaginary time, and their ratio therefore measures how long-lived the moment is relative to its size. For hydrogen this ratio is $\chi_S/\langle m^2\rangle \approx 0.032$ at tetrahedral sites but $0.046$--$0.050$ at octahedral sites: the octahedral moment is not merely
larger but also longer-lived per unit moment. As we show below, this
distinction is not incidental: octahedral and tetrahedral H sites occupy
geometrically distinct positions relative to the Lu $d$ orbitals, and this governs which $d$ channel each couples to. 

The spin and charge susceptibilities are moreover nearly equal on hydrogen ($\chi_S/\chi_D = 1.1$--$1.3$ at every site and composition), the signature of a site whose fluctuations are hybridization-driven rather than moment-forming, since local-moment behavior requires charge fluctuations to be suppressed relative to spin fluctuations. The $d$ shell behaves differently: $\chi_S/\chi_D$ falls from $4.6$--$5.4$ near the dihydride to $1.7$--$1.9$ at the trihydride, and $\chi_S$ itself collapses by a factor of four across the same range, tracking the emptying of the $d$ band. The $d$ manifold is thus moment-like — behaving as a localized, correlation-susceptible orbital — where it is close to half-filled, and becomes progressively itinerant or band-like as hydrogen withdraws its charge and the shell empties. Hydrogen, by contrast, remains itinerant and does not develop local-moment character at any composition, consistent with its H-$s$ band being governed by filling rather than by interaction strength. We note that $\chi_S$ is reported here at a single temperature ($\beta = 50$~eV$^{-1}$); distinguishing Curie from Pauli behavior would require its temperature dependence.

In the ideal fluorite structure the Lu site has $O_h$ symmetry, and each
hydrogen sublattice hybridizes with only one $d$ channel: tetrahedral H
lies along $\langle111\rangle$, where the $s$--$e_g$ two-center integrals
vanish, and octahedral H along $\langle100\rangle$, where the $s$--$t_{2g}$
integrals vanish~\cite{SlaterKoster1954}. Since the H-$1s$ level lies well
below $E_F$, the coupled $d$ set is pushed upward by antibonding repulsion
while the uncoupled set is left unshifted. In the tetrahedral-only
dihydride this raises $t_{2g}$ and leaves $e_g$ essentially nonbonding, so
charge accumulates preferentially in $e_g$: $n_{e_g} = 0.49$ against
$n_{t_{2g}} = 0.12$. (At intermediate compositions the octahedral
sublattice is partially occupied and the site symmetry is reduced, so the
separation is only approximate; it is exact in stoichiometric LuH$_2$ and
LuH$_3$.) Because the tetrahedral sublattice is occupied at every composition, the
$t_{2g}$ environment is fixed across the series; only the octahedral
sublattice varies, and it couples to $e_g$ alone. Adding additional hydrogen to the octahedral site therefore
acts almost exclusively on the $e_g$ channel, which acquires a hybridization
partner of its own and is pushed up in turn. The occupations bear this out:
from LuH$_2$ to LuH$_3$, $n_{e_g}$ falls by a factor of four
($0.49 \to 0.12$; $0.11$ in LaH$_3$) while $n_{t_{2g}}$ is nearly unchanged
($0.12 \to 0.11$), closing the imbalance from $n_{e_g}/n_{t_{2g}} \approx 4$
to $\approx 1.1$. Both sets remain far from the half-filled configuration
$n_{\rm orb} = 1$ throughout. The composition dependence of the $d$-shell
occupations thus follows from hybridization geometry and filling rather
than from any change in interaction strength.

The local spin susceptibility tracks the $e_g$ occupation across the series: the compositions with $n_{e_g} \gtrsim 0.36$ have $\chi_S = 0.16$--$0.22$, while those with $n_{e_g} \lesssim 0.19$ have $\chi_S = 0.05$--$0.08$. The $d$-shell spin fluctuations therefore follow
the filling of the channel that hydrogen depopulates, independently of
any change in the interaction parameters. The trends above follow from changing the hydrogen content, which alters
the filling and the hybridization geometry together. To separate the two we
consider rigid-band doping at fixed structure, which changes the carrier
count without touching either hydrogen sublattice (Table~S10).

The correlated-subspace occupations show where the doped charge is accommodated
(Table~S10). Across the full $\pm5\%$ range the Lu-$d$
occupation shifts by $0.51$ electrons, while both hydrogen sites change by less
than $0.005$---below the resolution at which we quote them. The local
susceptibilities behave the same way: $\chi_S$ on the $d$ shell rises by $55\%$
from hole- to electron-doping and its ratio to the charge susceptibility climbs
from $4.0$ to $6.0$, whereas the hydrogen susceptibilities and their ratios are
unchanged to within $3\%$ at both sites. Doping therefore acts exclusively on
the $d$ manifold; the hydrogen sublattices, including the more strongly hybridized octahedral
site, are spectators. The two perturbations agree: whether the $d$ filling is changed
chemically, by octahedral hydrogen, or rigidly, by doping, $\chi_S$ on the
$d$ shell follows the filling while the hydrogen sites do not respond.

The picture that emerges---a nearly filled but hybridized hydrogen shell,
weak renormalization in both channels, and an octahedral environment (both
the H site and the $e_g$ orbitals it couples to by symmetry) that is
systematically the more strongly hybridized of the two coordination
geometries---is one in which every site-resolved distinction traces back to
occupation and orbital geometry rather than to interaction strength. This
is the central claim of the present work in microscopic form, and it makes
a direct prediction for the spectral function: features appearing near
$E_F$ should be one-body in origin and controlled by octahedral occupancy
rather than by interaction strength. We test this against photoemission in
the following section.

\subsection{Electronic Structure, Correlated Spectral Function, and Optical Response }
\label{sec:results-experiments}

To trace the band-structure origin of the configurational dependence
observed in the photoemission and optical response (Figs.\ref{fig:akw},~\ref{fig:absorptivity_LuHx},), we computed the electronic band structures of four representative
LuH$_x$ configurations: (a) pure LuH$_2$ with full tetrahedral
occupancy, (b) LuH$_{1.75}$ with one tetrahedral H removed,
(c) LuH$_{2.25}$ with one additional octahedral H, and (d) nominally
LuH$_2$ with one tetrahedral vacancy compensated by one octahedral
occupant (see Fig. 4).

		\begin{figure*}[!ht]
	\centering
	\includegraphics[width=0.8\linewidth]{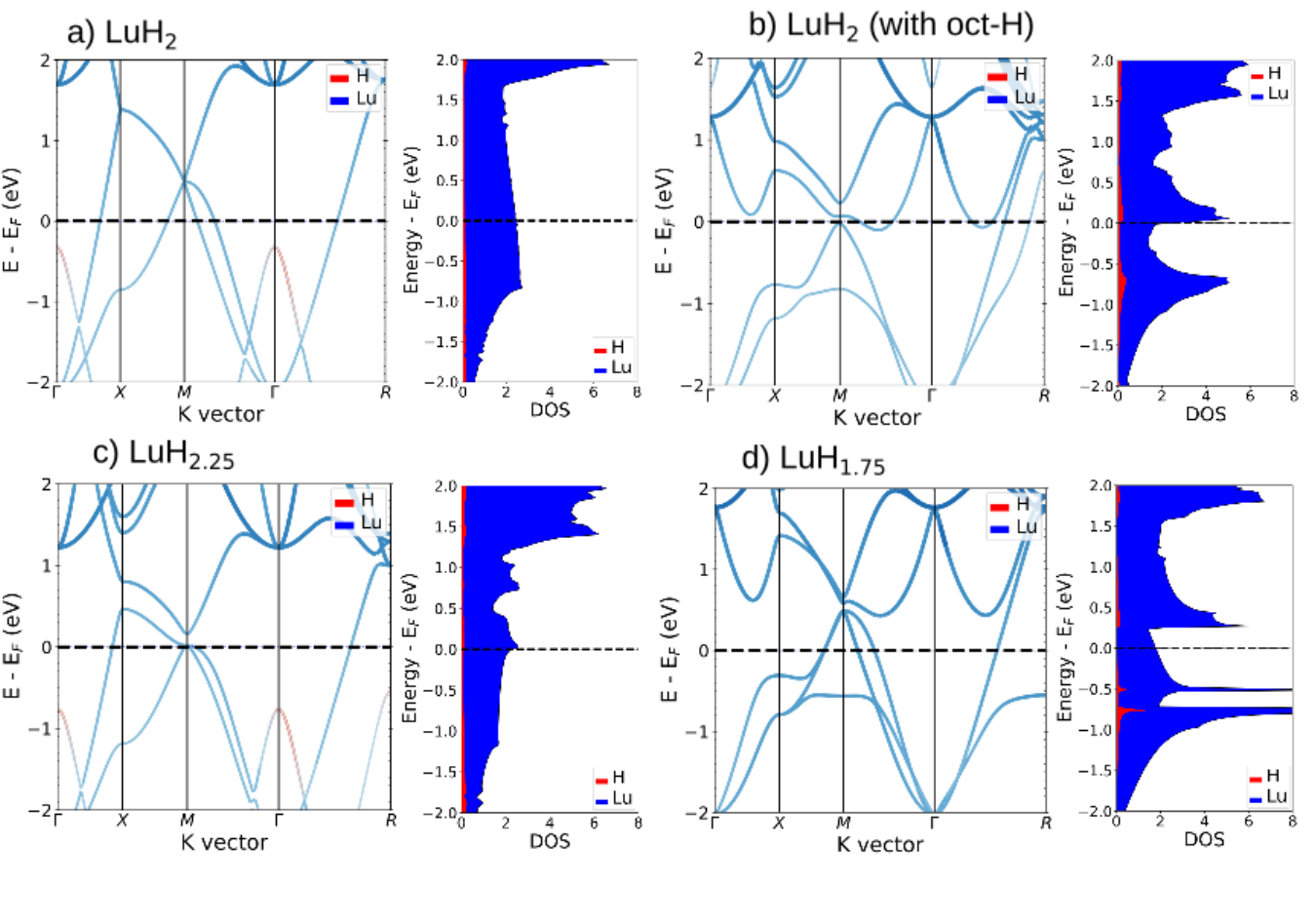}
	\caption{Comparision of band structures of various LuH$_x$ structures showing that flat-bands and van Hove singularity can be tuned near $E_F$. The band structures are for the following structures: (a) LuH$_2$ with only tetrahedral-H occupancy (b) LuH$_{1.75}$ with one tetrahedral-H vacancy (c) LuH$_{2.25}$ with one octahedral-H occupancy (d) LuH$_{2}$ with one tetrahedral-H vacancy and one octahedral-H occupancy.}
	\label{fig:vacancy-engineer}
\end{figure*}

The pure LuH$_2$ band structure (panel a) is the canonical
fluorite-dihydride result: dispersive Lu-5d bands crossing $E_F$
along $\Gamma-X-M-\Gamma-R$, with H-derived weight confined well below the
Fermi level. The two H-sublattice modifications produce distinct
changes near $E_F$. Removing a tetrahedral H (panel d, LuH$_{1.75}$)
introduces a substantially narrow band approximately $0.2$~eV below
$E_F$ along $M-\Gamma$.
Adding an octahedral H without removing a tetrahedral one (panel c,
LuH$_{2.25}$) shifts the band crossing at M downward onto $E_F$,
bringing a saddle-point-like feature into coincidence with the
Fermi level. The combined configuration (panel b) exhibits both
effects simultaneously: a weakly dispersive band near $E_F$
inherited from the tetrahedral vacancy, and a van Hove-like feature
at M shifted onto $E_F$ by the octahedral occupant.

The two effects therefore appear to act through different
mechanisms. Tetrahedral vacancies introduce a localised
defect-derived state whose narrow bandwidth reflects the loss of
metal-hydrogen hybridization in the immediate vacancy environment. Octahedral
occupants shift the Fermi level through electron donation while
simultaneously perturbing the Lu-5d states they hybridize with,
moving pre-existing dispersion features into coincidence with $E_F$.
The combination yields enhanced density of states
at the Fermi level relative to the pure dihydride.

The states introduced by tetrahedral vacancies and octahedral occupants are
narrow, and it is for narrow bands that correlation can least safely be assumed
away, $U/W$ being largest for precisely the weakly dispersing features these
modifications place near $E_F$ --- and the ARPES flat band has itself been
interpreted as a correlation feature~\cite{liang_observation_nodate}. Whether
such a band is correlated in the many-body sense cannot be decided from the band
structure alone, nor from a static Hubbard correction, which shifts levels but
cannot distinguish a band reconstructed by a static self-energy from one narrowed
by a small quasiparticle residue. We therefore turn to DMFT, which treats the
frequency dependence of the self-energy explicitly.

Correlation neither removes nor shifts these features. The flat band at $E_F$ and
the feature ${\approx}0.5$~eV below it survive the inclusion of dynamical
correlations essentially unchanged, so that both remain in agreement with the
ARPES measurements, complementing the agreement already obtained for the optical
conductivity. The self-energies underlying this conclusion, and their decomposition into static shift, mass renormalization and damping, are given in
Secion S4.

Figure~\ref{fig:akw} compares the Wannier bands of LuH$_2$ containing
octahedral hydrogen with the DMFT spectral function $A(\mathbf{k},\omega)$
along $X$--$\Gamma$--$M$. Because both are built from the same
Wannier Hamiltonian, every difference between them is attributable to the
self-energy. Three effects are visible, and we keep them separate throughout.

		\begin{figure*}[t]
	\centering
	\includegraphics[width=1.05\linewidth]{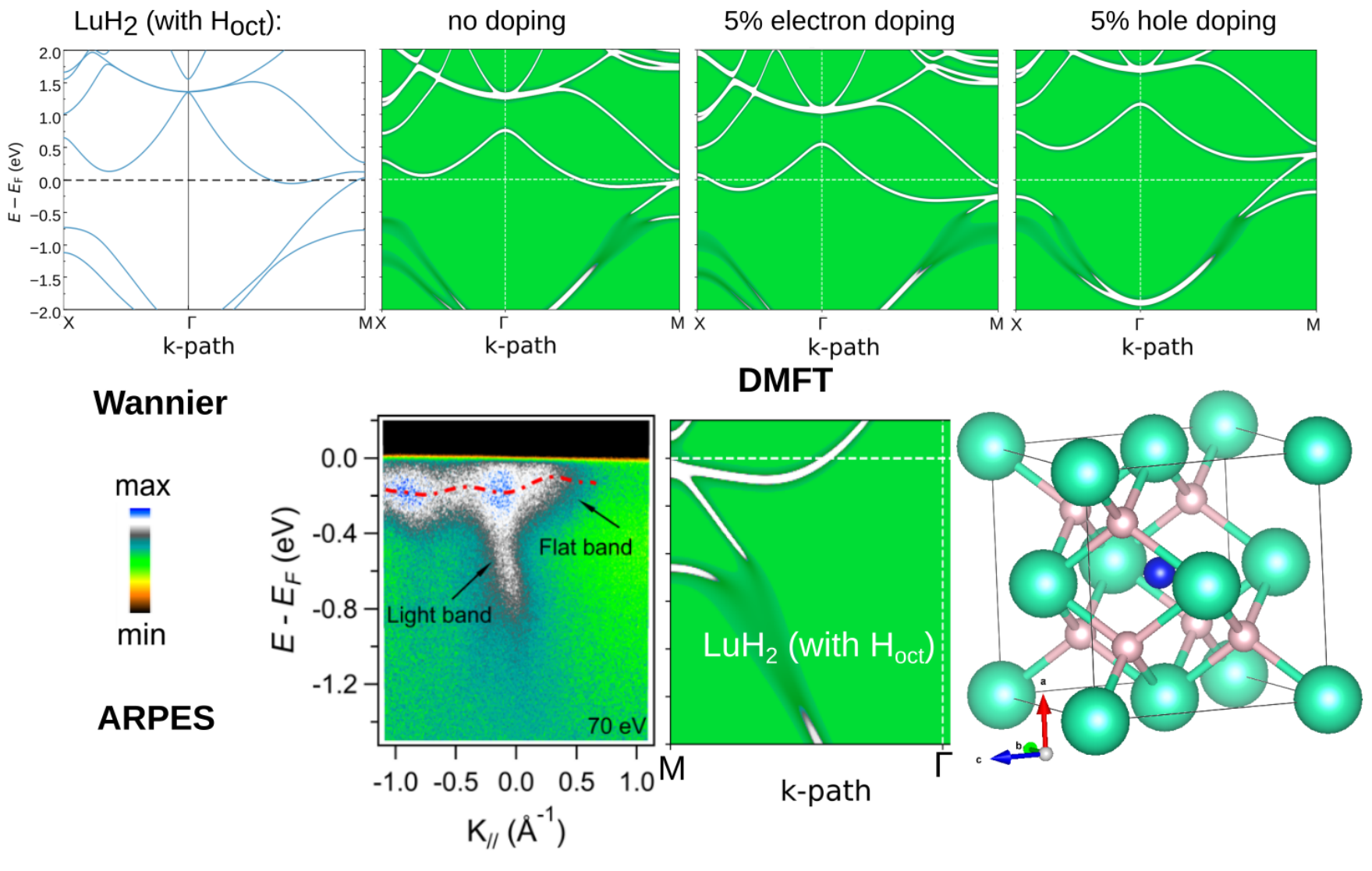}

	\caption{\textbf{Correlated electronic structure of LuH$_2$ with H$_{oct}$: theory--experiment comparison.}
		\textbf{Top:} Wannier bands and DMFT spectral function $A(\mathbf{k},\omega)$
		along $\Gamma$--$X$--$M$, undoped and at $\pm5\%$ carrier
		doping. Relative to the Wannier bands, DMFT produces a narrow band pinned
		near $E_F$ with a saddle at $M$, and damped flat features near $-0.6$~eV;
		doping shifts both rigidly in energy (electron doping downward, hole doping
		upward) without displacing them in momentum.
		\textbf{Bottom:} ARPES cuts adapted from Ref.~\cite{liang_observation_nodate}, licensed under CC by 4.0, showing
		the near-$E_F$ flat band
		with its emerging light band (left), compared with the undoped DMFT
		spectral function along $M$--$\Gamma$ (middle) and the model structure of LuH$_2$ with H$_{oct}$ -- teal atoms represent Lu, pink atoms represent H$_{tet}$ and blue atom represents H$_{oct}$(right).
	}

	\label{fig:akw}
\end{figure*}

The first and largest is static. The orbital-dependent 
$\mathrm{Re}\,\Sigma_\alpha(0)-V^{\rm dc}$ repositions the
hydrogen-derived states relative to the Lu-$d$ manifold, and where the
resulting weakly dispersing level intersects bands of predominantly Lu-$d$
character, hybridization quenches their dispersion. The clearest consequence
is along $\Gamma$--$X$--$M$: a band that disperses appreciably in the Wannier
spectrum becomes a narrow, nearly flat feature pinned close to the Fermi
level. At $M$ this narrow band forms a saddle, hole-like along one in-plane
direction and electron-like along the other, lying within a few tens of meV
of $E_F$.

The second effect, dynamical mass renormalization, is comparatively weak. The
Matsubara self-energies give $m^*/m\lesssim1.36$ for either of $d$ or $s$ orbitals, bounding the correlation-induced
bandwidth reduction to at most $\sim$36\%. The observed flattening of the
near-$E_F$ band substantially exceeds this bound, identifying it as a
reconstruction of the band structure by the static, orbital-dependent part of
the self-energy rather than as a mass-enhancement effect.

The third effect is damping. A second set of flat features appears near
$-0.6$~eV, but with markedly lower intensity and larger energy width than the
near-$E_F$ band. These features are flat in the dispersive sense; what
distinguishes them is lifetime, since $|\mathrm{Im}\,\Sigma(\omega)|$ grows
away from the Fermi level and broadens them into low-intensity ridges. The
narrow band near $E_F$ remains sharp and intense by contrast, protected by the
vanishing of $\mathrm{Im}\,\Sigma$ at low frequency.

The $\pm5\%$ doped spectra differ from the stoichiometric one by an
essentially rigid displacement in energy. With each spectrum referenced to its
own Fermi level, electron doping raises the chemical potential and the narrow
band and its saddle appear at lower energy, moving further below $E_F$;
hole doping shifts them upward toward and through $E_F$. No displacement in
momentum accompanies these shifts, as required for a momentum-independent
self-energy, and the dispersions, intensities and linewidths are unchanged
within our resolution. Linear extrapolation of the saddle energy places it at
the Fermi level for a hole doping of $5\%$.

For the LuH$_2$ with H$_{oct}$ structure we also calculate the in-plane and out of plane energy distribution curves (EDC) and compare with EDC curves obtained from ARPES experiment (Figure S3). \cite{liang_observation_nodate}. The construction of such curves, which introduces no adjustable parameters beyond the experimental
temperature, energy resolution, and escape-depth momentum broadening, is
described in Section S4.3. In plane, a sharp maximum sits immediately below $E_F$ at fixed energy across the
cut, with deeper structure that shifts markedly from curve to curve---the
distinction drawn in the measured curves between a flat band and the light band
emerging from it, with the calculated near-$E_F$ maximum at $-50$~meV against a
measured flat-band binding energy of $0$--$0.2$~eV. Out of plane, both show a
stationary maximum just below $E_F$, persisting across photon energies spanning
more than one Brillouin zone along $k_z$; in the calculation this peak is
stationary to within $10$~meV over the full range. The calculation
additionally resolves a second out-of-plane maximum near $-0.6$~eV, dispersing
by $0.15$~eV, which appears in the measurement not as a distinct peak but
within the broad enhancement between $-0.5$ and $-0.9$~eV; we attribute this to
momentum integration over the escape-depth window and to the ordered octahedral
sublattice of our supercell, which produces coherent minibands where a
disordered sample would show an incoherent resonance at the same energy
(Sec.~S4.3). The comparison is therefore one of feature count, approximate
energy, and dispersive character; peak intensities are not compared, since the
measured curves carry photoemission matrix elements and an inelastic background
absent from $A(\mathbf{k},\omega)$. Likewise, the calculated DMFT DOS $N(\omega)$ for LuH$_2$ with H$_{oct}$
exhibits a pronounced peak at the Fermi level and a broad structure centred
near $-0.6$~eV, separated by a shallow minimum (Figure S4). The Fermi-level peak is the density-of-states signature of the narrow band identified above: a band of
width $W$ concentrates its spectral weight over that width, so a band of
$\sim$0.2~eV width at $E_F$ necessarily produces a peak of this kind. Its
orbital decomposition is dominated by Lu-$d$ states, with the $e_g$ set
carrying the larger share and hydrogen $s$ weight negligible in this window,
the latter having been displaced to higher binding energy by the static shift.

Taken together, these results show that the correlated orbital creates a flat
band it does not itself occupy. Correlation on the hydrogen $s$ states is
substantial but almost entirely static:
$\mathrm{Re}\,\Sigma_{\mathrm{H}\text{-}s}(0)$ displaces the H-derived weight to
higher binding energy, while $m^*/m \lesssim 1.3$ leaves the quasiparticle
residue close to unity. The narrow band remaining at $E_F$ is therefore Lu-$d$
in character, flattened where the displaced H-$s$ level crosses it---a
hybridization effect driven by a correlation-induced shift, not a heavy
quasiparticle. The same self-energy accounts for both photoemission features,
the sharp Lu-$d$ peak at $E_F$ protected by the vanishing of
$\mathrm{Im}\,\Sigma$ and the broad, damped H-$s$ structure near $-0.6$~eV.
Agreement with the measured flat-band binding energy improves under electron
doping. Because the mechanism is a level alignment rather than a mass
enhancement, its strength is set by where the hydrogen sits and how much of it
is present---which is what the comparisons that follow address.

These single-particle differences appear directly in the optical
response. Figure~\ref{fig:absorptivity_LuHx} shows the absorptivity
$A(\omega) = 1 - R(\omega)$, computed from the independent-particle
dielectric function (Fig.~S4), for the tetrahedral-only series
[panel (b)] and for the octahedrally compensated LuH$_2$
[panel (a)], together with the measured spectra of Weaver
\emph{et al.}~\cite{Weaver1979optical}. Configurations (a) and (d)
share the same stoichiometry, so any difference between their
spectra is structural rather than compositional.

In pure LuH$_2$, $\varepsilon_2$ falls below unity between $1.0$ and
$2.2$~eV (Figure ~S4): the tetrahedral-only band structure supports no
direct transitions in this window, hydrogen weight lying several eV
below $E_F$. Within the window $\varepsilon_1$ crosses zero at
$1.05$~eV with $\varepsilon_2$ of order unity, so the loss function
$-\mathrm{Im}\,\varepsilon^{-1}$ reaches order unity and the
absorptivity shows the corresponding sharp plasma edge --- a single
maximum of $0.96$ at $1.05$~eV over an essentially flat sub-edge
background, the canonical Drude-plus-interband response of a clean
CaF$_2$-structure dihydride. Tetrahedral vacancies (LuH$_{1.875}$,
LuH$_{1.75}$) progressively broaden this edge, raise the low-energy
background, and shift the maximum toward $1.7$--$2.0$~eV, but
introduce no new structure below it.

Octahedral occupancy does. The compensated structure develops an
occupied H-derived peak near $-0.65$~eV in the projected density of
states (Figure ~S2), with no counterpart in any tetrahedral-only
configuration. Transitions from this state into the unoccupied
Lu-5$d$ manifold set in near $0.65$~eV and produce an
$\varepsilon_2$ oscillator peaking at $1.05$~eV, consistent with the
unoccupied DOS maximum at $\approx +0.4$~eV. The zero of
$\varepsilon_1$ moves down to $0.57$~eV, where now
$\varepsilon_2 \approx 15$ and the loss function is reduced to
$\approx 0.07$; the single sharp maximum is replaced by two damped
ones, at $0.7$ and $1.75$~eV, separated by a minimum near
$1.15$~eV, and the integrated spectral weight below $1$~eV
substantially exceeds that of any tetrahedral-only spectrum. Both
maxima fall at minima of $\varepsilon_2$ rather than at its peaks:
for $|\varepsilon| \gg 1$ the reflectivity tracks $|\varepsilon|$,
so $A$ is largest where $|\varepsilon|$ is smallest, and neither
maximum marks a transition energy.

	\section{Discussion}

We now examine the physical origin of the trends established above. We first consider how the hydrogen on-site energies fix the sublattice occupations, and how the resulting filling — together with the screening environment set by the Lu-d manifold — determines the correlation strength on hydrogen. We then compare our results with optical and photoemission measurements on lutetium and related lanthanide hydrides, and assess what they imply for the electronic character of the flat bands near the Fermi level.

\subsection{Origin of Filling and Correlation}
\label{sec:disc-onsite}

The on-site energies provide a direct, one-body explanation for the hydrogen
filling that underlies our central result. An orbital's occupation is set by its
energy relative to the Fermi level, and in LuH$_3$ both H-$s$ sublattices lie
$3$--$4$~eV below $E_F$: the hydrogen shells are therefore filled and anionic
(H$^-$-like), consistent with the near-full DMFT occupation
($n_{\mathrm{H}\text{-}s}\approx1.75$) and with photoemission evidence for
charge transfer to hydrogen~\cite{Weaver1979pes,VandeWalle2003}.

The Wannier analysis further reveals the electronic inequivalence of the two
hydrogen sublattices: the octahedral H-$s$ level lies $\sim1$~eV above the
tetrahedral one, identifying the octahedral hydrogen as the more weakly bound,
shallower species. This is significant because the octahedral hydrogen is the
common actor across the phenomenology of these materials: it controls the
carrier density probed optically~\cite{Weaver1979optical}, it is the Kondo-like
impurity of the breathing-hydrogen picture~\cite{Eder1997,Ng1999}, and it is the
sublattice whose distortion opens the gap in cubic
LuH$_3$~\cite{Denchfield2025}. Our calculation shows directly that, for lutetium hydrides, it is the
hydrogen level closest to $E_F$---the least deeply bound and hence the most
likely to be electronically and structurally active---while the tetrahedral
hydrogen, filled already in the dihydride, lies deeper and is correspondingly
inert.

The susceptibilities sharpen this picture and provide the most robust
site-resolved discriminant in our data. In every Lu composition the octahedral
hydrogen carries $1.8$--$2.8$ times the local spin susceptibility of the
tetrahedral hydrogen, without exception---including LuH$_{2.875}$, where the
mass enhancements of the two sites are inverted relative to the other
compositions. Because $\chi_S$ is a static thermodynamic average rather than a
low-frequency derivative of a stochastic quantity, it is the less noise-sensitive
diagnostic, and we take the susceptibility ordering as the more reliable
statement of site inequivalence. Its ordering follows the on-site energies
directly: the shallower octahedral level supports larger and longer-lived local
fluctuations than the deeper tetrahedral one. The reversal in LaH$_3$, where the
octahedral level lies below the tetrahedral one, is reproduced in the
susceptibilities as well---octahedral $\chi_S$ falls below tetrahedral---so that
the level ordering, the occupations, the local moments, and the susceptibilities
all identify the same sublattice as the electronically active one in each
compound. That four independent diagnostics agree indicates the site hierarchy
is a robust feature of the electronic structure rather than an artifact of any
single observable.

The level alignment directly rationalizes the weak hydrogen correlation. A shell
buried well below $E_F$ is filled and far from the half-filled configuration
where the interaction could localize it; the large H-$s$ interaction therefore
acts only as a static level shift rather than a mass renormalization, exactly as
found in our DMFT self-energies. Level position, filling, and correlation thus
form a single causal chain: the host sets the hydrogen level, the level sets the
filling, and the filling---not the magnitude of $U$---controls the correlation.
This is the filling-control principle familiar from Hund's metals and
orbital-selective correlation in $d$-electron
systems~\cite{Georges2013,Mravlje2011}, operating here in its
simplest setting: a single $s$ orbital with negligible Hund's coupling. The same
principle couples the two manifolds---the hydrogen acts as the electron
reservoir that sets the $d$-band filling and thereby its (weak) correlation---as
detailed in Section~3C.

The rigid-band doping calculations make the reservoir behavior explicit. Adding
or removing one electron per formula unit shifts the Lu-$d$ occupation by the
full doped charge while the hydrogen occupations, moments, and local
susceptibilities are unchanged---at the octahedral site as well as the
tetrahedral one, even though the octahedral hydrogen is the more electronically
active of the two. The hydrogen shell is thus closed not only in its static
occupation but in its response: charge introduced at the Fermi level is absorbed
entirely by the $d$ manifold. This is the dynamical counterpart of the deep,
filled anionic level identified in the on-site energies, and it is the condition
under which the rigid-band comparison with photoemission
(Secion~\ref{sec:results-experiments}) is justified---the hydrogen-derived states near
$E_F$ shift with the chemical potential but do not themselves reconstruct. The
$d$ susceptibility, by contrast, grows steadily with filling and becomes
increasingly spin-dominated ($\chi_S/\chi_D$ rising from $4.0$ to $6.0$), the
same trend seen across the stoichiometric series as hydrogen content is reduced.
The two routes---varying hydrogen content at fixed electron count, and varying
electron count at fixed hydrogen content---thus give a consistent account of
which manifold is compressible.

The $d$-channel behavior follows from the crystal-field structure of the
fluorite lattice. The metal is coordinated by eight tetrahedral hydrogen at the
cube corners rather than by six ligands octahedrally, and the cubic field inverts the familiar ordering: $e_g$, whose lobes point between the ligands, lies below $t_{2g}$. The $e_g$ set consequently fills first and sits closer to half filling, carrying the larger per-orbital moment. Adding octahedral hydrogen introduces a competing field of the opposite sign---the $4b$ site \emph{is}
octahedrally disposed with respect to the metal---which partially cancels the
cubic splitting; this accounts for the narrowing of the $e_g$/$t_{2g}$
occupation contrast from the dihydride to the trihydride.

Because both sets lie below half filling throughout, the spin susceptibility is
on the rising side of its filling dependence, and the $d$ shell moves
\emph{away} from the moment-forming regime as hydrogen content increases. The rigid-band doping calculations confirm this directly: adding electrons raises $n_d$ and $\chi_S$ monotonically with no turnover (Table~18), as expected when approaching half filling from below. The stoichiometric series and the doping series therefore describe the same dependence traversed in opposite directions, and identify the $e_g$ occupation---set by hydrogen content---as the variable controlling the $d$-shell spin response.

\subsection{Comparision with Experiment}
\label{sec:disc-experiment}

Angle-resolved photoemission on (nominally nitrogen-doped) lutetium hydride
reports a flat band $0$--$200$~meV below $E_F$ and a nearby van Hove
singularity, interpreted as a signature of strong correlation~\cite{liang_observation_nodate}.
Our results bear directly on the origin of this feature. The DMFT Lu-$5d$ mass
enhancement is modest ($m^*/m \approx 1.1$--$1.3$), whereas a
correlation-driven flat band would require a heavy quasiparticle mass,
$Z \ll 1$. We therefore find that the near-$E_F$ flat band is unlikely to be
primarily a correlation-renormalization effect, pointing instead toward a
disorder-related (hydrogen-vacancy) or filling/Van Hove origin, consistent with
the absence of such a band in pure-LuH$_2$ DFT~\cite{liang_observation_nodate}. We note that the
precise hydrogen and nitrogen content of the ARPES sample is not firmly
established; if the sample is closer to LuH$_{2-x}$ than to a nitrogen-doped
composition, this conclusion is unaffected, since the argument rests on the
magnitude of the computed renormalization rather than on the dopant.

\begin{figure}[t]
	\centering
	\includegraphics[width=1.05\linewidth]{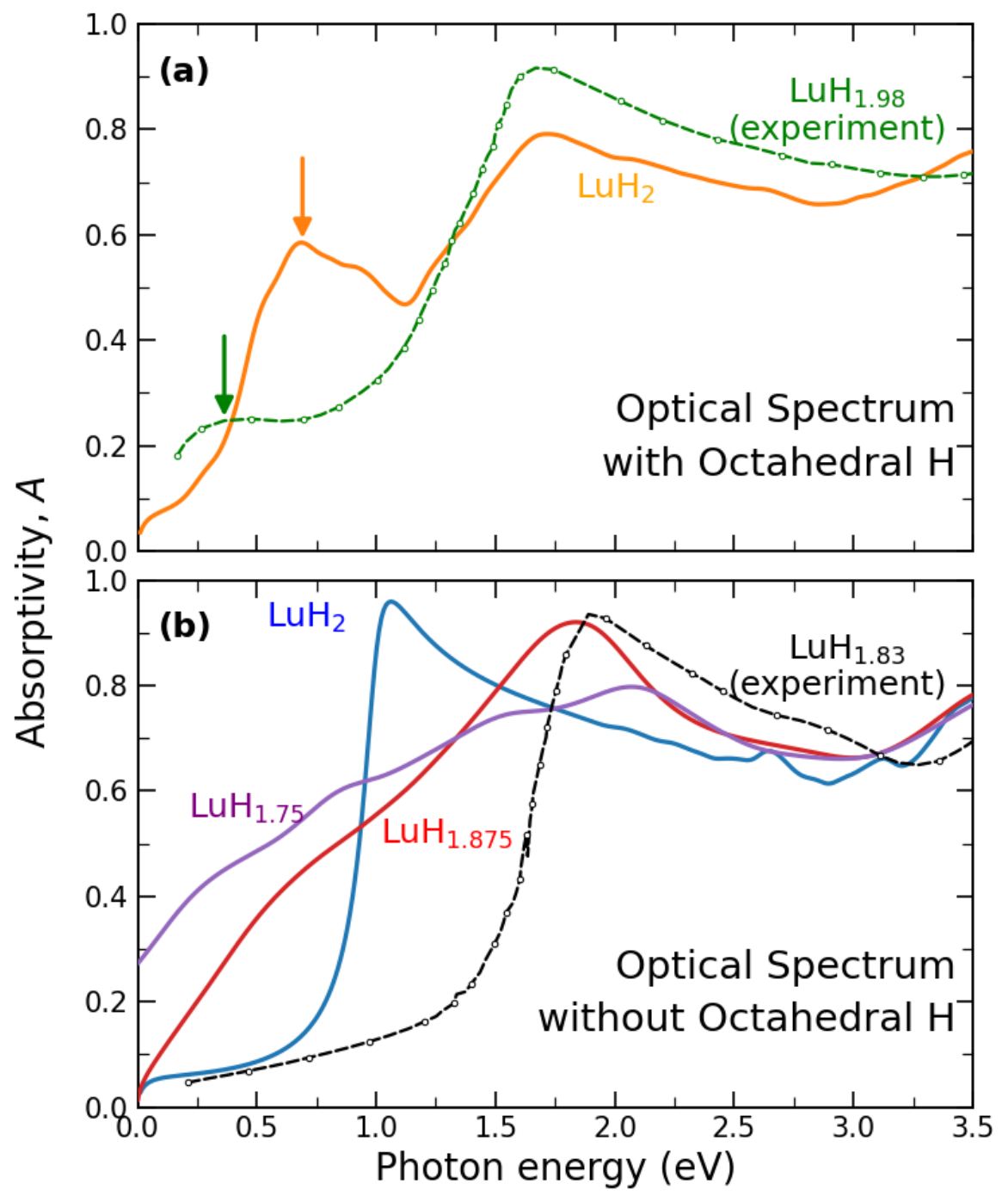}
	\caption{Calculated optical absorption spectra of LuH$_x$ structures compared with
		experiment. 
		(a) Calculated absorptivity for a LuH$_2$ structure in which one hydrogen
		occupies an octahedral site, compared with experimental data for LuH$_{1.98}$,
		which also shows weak octahedral occupation. Arrows mark the low-energy feature
		associated with transitions involving octahedral hydrogen: orange for the
		calculated LuH$_2$ curve and green for the experimental LuH$_{1.98}$ curve.
		This feature is absent in the tetrahedral-only structures of panel (b). (b) Calculated absorptivity for LuH$_2$, LuH$_{1.875}$, and
		LuH$_{1.75}$, which contain only tetrahedral hydrogen, compared with
		experimental data for LuH$_{1.83}$, which is likewise tetrahedral-only.}
	\label{fig:absorptivity_LuHx}
\end{figure}

Figure~\ref{fig:absorptivity_LuHx} compares the calculated optical
absorptivity $A(\omega)$ of LuH$_x$ in four configurations against the
experimental spectra of Weaver \emph{et al.}~\cite{Weaver1979optical} at
$x=1.83$ and $x=1.98$. The dielectric functions (Fig.~S[E]) show why octahedral occupation,
and only octahedral occupation, restructures the spectrum. The
oscillator introduced by the octahedral H falls inside the
transparency window of pure LuH$_2$ --- the same window that hosts
its plasma edge. Filling the window has two consequences: the added
interband screening shifts the zero of $\varepsilon_1$ from $1.05$
down to $0.57$~eV, and $\varepsilon_2$ at the crossing rises by an
order of magnitude, collapsing the loss function. The sharp plasma
edge of the dihydride is thereby eliminated, leaving a damped
feature at $0.7$~eV and a residual maximum at $1.75$~eV in the part
of the window that survives above the oscillator. Tetrahedral
vacancies, by contrast, redistribute oscillator strength within the
existing bands: they broaden the edge, but the window --- and with
it the single-maximum line shape --- remains.

The measured spectra display the same dichotomy. LuH$_{1.83}$ shows
a sharp edge and no sub-eV structure, consistent with predominantly
tetrahedral occupation near the dihydride phase boundary; its
maximum at $2.05$~eV is bracketed by the calculated LuH$_2$
($1.05$~eV) and LuH$_{1.875}$ ($1.85$~eV). Near-stoichiometric
LuH$_{1.98}$ shows an additional shoulder at $0.3$--$0.5$~eV absent
from LuH$_{1.83}$, in the region where the octahedral calculation
peaks, though at somewhat lower energy and with far smaller
amplitude: $\Delta A \approx 0.02$ against the calculated
$\approx 0.12$, so any octahedral occupancy in the measured sample
must be small. Two caveats attend the assignment. A Drude term with
a finite scattering rate produces an inflection of similar shape
with no interband contribution at all, and a firm identification
would require the calculated absorptivity followed as a continuous
function of octahedral filling, together with an independent
structural determination of the site occupancies. A systematic
offset of the calculated edge relative to experiment is also
apparent --- $1.05$~eV for the all-tet reference against the
measured $\sim 1.7$~eV; we discuss its possible origins in
Sec.~\ref{sec:octahedral_discussion}, and the interpretation here
rests on the relative behavior across configurations, which is
reproduced unambiguously.

The configurational comparison in Fig.~\ref{fig:absorptivity_LuHx}
provides a direct band-structure-level confirmation of the inference
first made by Weaver \emph{et al.}~\cite{Weaver1979optical} that the sub-eV
optical feature observed at near-stoichiometric compositions arises
from octahedral H occupation rather than from tetrahedral disorder.
The crucial control is the comparison between LuH$_{1.75}$ and
LuH$_{1.875}$ (both tetrahedral-only) and LuH$_2$ with a single
octahedral occupant. Tetrahedral vacancies smear the plasmon edge but
do not introduce a new oscillator anywhere below 1.5~eV; octahedral
occupation does. This separates site type from H content as the
controlling variable---a separation that the original optical study
could only approach indirectly through the quench-versus-slow-cool
experiment on YH$_{1.89}$ and through the systematic lattice-constant
trend across Sc, Lu, Y.

It is worth recalling how the octahedral assignment was established
historically, since the present calculation completes a chain of
reasoning begun nearly half a century ago. Weaver \emph{et al.}'s
1979 optical inference~\cite{Weaver1979optical} was supported on the
theoretical side by the companion self-consistent band calculation
of Peterman \emph{et al.}~\cite{PhysRevB.19.4867}, which showed that
tetrahedral-only CaF$_2$-structure YH$_2$ has no interband
transitions in the $0.5$--$1.5$~eV window, establishing that the
sub-eV optical features required states outside the tetrahedral
picture. Direct structural confirmation came shortly afterwards:
proton NMR rigid-lattice second moments found $10$--$15\%$
octahedral occupation in YH$_{1.92}$ and
YH$_{1.98}$~\cite{Khatamian1980}, and neutron diffraction and
inelastic scattering subsequently tracked the temperature dependence
of the octahedral occupancy in YH$_{2.0}$ and identified its
optic-mode signature~\cite{Anderson1982}. For LuH$_x$ specifically
the direct structural evidence is sparser, and the present supercell
calculation occupies a role analogous to Peterman's but in the
affirmative: where the tetrahedral-only band structure could only
\emph{exclude} certain transitions, we \emph{include} an octahedral
occupant and show spectroscopically that it produces the observed
feature.

The dichotomy between tetrahedral and octahedral hydrogen is rooted in local
coordination. In the fluorite structure, tetrahedrally coordinated H sits in a
four-fold metal environment with comparatively short metal--hydrogen distances
and strong $d$--$s$ hybridization, which pulls the tetrahedral bonding band well
below $E_F$. Octahedrally coordinated H has six metal neighbors at longer
distance and weaker hybridization, placing its bonding combination closer to
$E_F$. This ordering---tetrahedral deep, octahedral shallow---is the one we
obtain for LuH$_3$, and it is the ordering established in the early rare-earth
trihydride band-structure
literature~\cite{Switendick1970,Gupta1982,MisemerHarmon1982}. Both sites lie
well below the Fermi level, consistent with the universal H$^-$ level-alignment
framework of Van de Walle and Neugebauer~\cite{VandeWalle2003}.

Coordination is not, however, the only variable. The octahedral interstitial is
the larger and less tightly confined of the two sites, and its level is
correspondingly more sensitive to the cell volume. In LaH$_3$, whose lattice
constant exceeds that of LuH$_3$ by more than $10\%$, this sensitivity is
sufficient to reverse the ordering, placing the octahedral level below the
tetrahedral one (Table~\ref{tab:onsite}). The tet/oct hierarchy therefore
reflects a competition between coordination-driven hybridization, which favors a
deeper tetrahedral level, and the volume dependence of the octahedral site; it
is not a fixed property of the fluorite structure.

This level alignment is precisely the hierarchy revealed by our
on-site Wannier energies in LuH$_3$, where the tetrahedral H~$1s$
level sits at $\varepsilon - E_F = -3.8$~eV and the octahedral
H~$1s$ level at $-2.9$~eV --- a $0.9$~eV splitting in the expected
direction (Sec.~\ref{sec:results-onsite}). The on-site energies set
the hierarchy, not the transition energies: hybridization with the
Lu-5$d$ manifold carries octahedral-derived weight up to the
occupied peak at $-0.65$~eV seen in the supercell density of states
(Fig.~S2), and it is transitions from this hybridized manifold into
the unoccupied Lu-5$d$ states that produce the oscillator at
$1.05$~eV. Tetrahedral-derived weight remains several eV below
$E_F$ even after hybridization, so its transitions lie far outside
the Weaver window. The sub-eV structure in our supercell is
therefore a direct spectroscopic signature of the shallow octahedral
manifold, not a generic stoichiometry effect.

The systematic dependence on the host metal across the trivalent
dihydrides reinforces this picture. The octahedral hole grows with
the lattice constant in the sequence ScH$_x$ ($a_0 \approx 4.78$~\AA),
LuH$_x$ (5.03~\AA), YH$_x$ (5.20~\AA), and the spectroscopic signature
of octahedral occupation follows in lockstep: absent in ScH$_x$,
weak near 0.4~eV in LuH$_x$, and strong with two distinct features at
0.35 and 1.25~eV in YH$_x$~\cite{Weaver1979optical}. This is consistent both
with the steric argument (larger octahedral holes are easier to
occupy) and with the bonding argument (the octahedral H bonding
level rises further toward $E_F$ as the metal-hydrogen octahedral distance
lengthens, increasing oscillator strength for transitions into the
conduction manifold).

This optical interpretation closes the loop on the picture assembled
from our cRPA, on-site-alignment, and mass-enhancement analyses,
which together identify the octahedral hydrogen sublattice as the
electronically active component of LuH$_x$ in the near-stoichiometric
regime. The screened interaction
$U_{\text{H-}s}^{\text{oct}}$ is systematically lower than
$U_{\text{H-}s}^{\text{tet}}$ in the H-$s$-only subspace and inverts
upon enlargement of the active space to include Lu-$d$ screening,
identifying the octahedral H as the more strongly coupled to the
metallic conduction electrons; the on-site Wannier energy places it
within hybridization range of the Fermi level; and the optical
response now supplies an independent, experimentally anchored
confirmation that the spectroscopic fingerprint of
stoichiometry-driven changes in LuH$_x$ near $x = 2$ resides in
transitions involving this shallow octahedral manifold.

\section{Conclusions}

We investigate electronic correlation in the rare-earth hydrides LuH$_x$ ($x = 1.75$--$3$) together with other LnH$_x$ compound using dynamical mean-field theory (DMFT) with the Hubbard U values obtained from constrained random-phase approximation 
(cRPA). Our studies reveal that 
hydrogen-derived $s$ orbitals has much larger screened on-site Coulomb interactions than the Lu $d$ states. A joint analysis of Wannier spread functions and interaction parameters (see discussion in Section S2.2) demonstrates that these trends originate from the interplay of orbital localization and screening efficiency as evidenced by the pronounced dependence of $U$ on the choice of correlated subspace.  The screened hydrogen interaction rises with hydrogen
content at nearly fixed lattice constant, driven by the progressive anionic
conversion of the octahedral hydrogen, which depletes the conduction-electron
screening; this is the same carrier-density reduction observed experimentally as
the decreasing Drude plasma frequency across the series~\cite{Weaver1979optical}, here manifested in the interaction parameters.

Despite carrying the largest interaction in the system, the hydrogen is weakly correlated. In hydrides LuH$_x$ ($x = 1.75$--$3$) and LaH$_3$, DMFT calculations indicates that these materials are weakly correlated metals; within this weakly correlated regime, however, the correlation is strongly filling-controlled. The LuH$_x$ hydrogen is a nearly filled, closed-shell anion with a deep on-site level consistent
with the photoemission evidence for charge transfer to hydrogen~\cite{Weaver1979pes}, whose large interaction is inert because the
filled shell admits no charge fluctuations. The Lu-$d$ band correlation in LuH$_x$, decreases as octahedral hydrogen dilutes the $d$ filling. The hydrogen content thus emerges as the principal tuning parameter for correlation in these materials---it is the electron reservoir that sets the filling of every manifold.

These results bear directly on the experimental phenomenology. Our calculations indicate that the flat bands
reported near $E_F$ by photoemission~\cite{liang_observation_nodate} arise from the presence of H$_{oct}$ and H$_{tet}$ vacancies. At the same time,
tetrahedral vacancies and octahedral occupancy can be tuned to place narrow
bands and van Hove singularities near $E_F$, so that hydrogen stoichiometry
offers a structural route to engineering the low-energy density of
states. The ~0.9 eV splitting we compute between the tetrahedral and
octahedral hydrogen levels, and the correlation contrast we predict between the
lutetium and lanthanum trihydrides, are in principle accessible to photoemission
and optical spectroscopy.

The broader lesson is a simple organizing principle for correlation in
hydrogen-rich materials: the host band structure sets the hydrogen level, the
level sets the filling, and the filling---not the magnitude of $U$---controls
the correlation. A site-resolved treatment distinguishing the two hydrogen
sublattices in LaH$_3$, where the hydrogen is correlation-active, is the natural
next step and would test the site-selective correlation this principle predicts.

\section{Acknowledgments}
This research was supported by the National Science Foundation (NSF)
(DMR-2104881, A.L., R.J.H.), the Department of Energy (DOE) National Nuclear
Security Administration (NNSA) through the Chicago/DOE Alliance Center
(DE-NA0004153; R.J.H.), the DOE Office of Science (SC) (DE-C0020340, R.J.H.),
and NSF SI2-SSE Grant 2513657 (H.P.). P. G. (mentorship, analysis, writing) and
H. S. (mentorship, analysis, writing) were supported by the DOE SC, Basic
Energy Sciences (BES), Materials Sciences and Engineering Division, as part of
the Computational Materials Sciences Program and Center for Predictive
Simulation of Functional Materials. The authors gratefully acknowledge the
Advanced Cyberinfrastructure for Education and Research (ACER) group at the
University of Illinois Chicago for providing computational resources and
services needed to deliver the research results presented in this paper.
Similarly, this research used resources of the National Energy Research
Scientific Computing Center (NERSC), a DOE SC User Facility located at
Lawrence Berkeley National Laboratory, operated under Contract No.
DE-AC02-05CH11231 using NERSC award BES-ERCAP0023615. This research used resources of the Argonne Leadership Computing Facility, which is a DOE SC User Facility supported under contract DE-AC02-06CH11357. This manuscript has been
authored by UT-Battelle, LLC under Contract No. DE-AC05-00OR22725 with the
DOE. Part of the DFT-based research was conducted as part of a user project at
the Center for Nanophase Materials Sciences (CNMS), a DOE SC User Facility at
Oak Ridge National Laboratory. Contributions to the manuscript were also
sponsored by UT-Battelle, LLC under the same contract. The U.S. Government
(USG) retains, and the publisher, by accepting the article for publication,
acknowledges that the USG retains, a non-exclusive, paid-up, irrevocable,
worldwide license to publish or reproduce the published form of this
manuscript, or allow others to do so, for USG purposes. The DOE will provide
public access to the results of federally sponsored research in accordance
with the DOE Public Access Plan
(http://energy.gov/downloads/doe-public-access-plan).

\bibliographystyle{apsrev4-2}
\bibliography{references}



\section*{Supplementary material}

# Origin of Flat Bands and Role of Electron Correlation in Lutetium Hydrides

Anmol Lamichhane*[1], Adam Denchfield[2], Hyeondeok Shin[3], Panchapakesan Ganesh[4], Russell J. Hemley[1,5,6], and Hyowon Park[1,7]

[1]Department of Physics, University of Illinois Chicago, Chicago, IL
[2]Oden Institute for Computational Engineering and Sciences, The University of Texas at Austin, Austin, Texas 78712, USA
[3]Computational Science Division, Argonne National Laboratory, Lemont, Illinois 60439, USA
[4]Center for Nanophase Materials Sciences, Oak Ridge National Laboratory, Oak Ridge, TN
[5]Department of Chemistry, University of Illinois Chicago, Chicago, IL
[6]Department of Earth and Environmental Sciences, University of Illinois Chicago, Chicago, IL
[7]Materials Science Division, Argonne National Laboratory, Lemont, Illinois 60439, USA

## 1 Computational details

First-principles calculations were performed within the framework of density functional theory (DFT) using the Vienna *Ab initio* Simulation Package (VASP) [4]. The projector augmented-wave (PAW) method [2] was employed to describe the interaction between valence electrons and ionic cores, and the exchange–correlation effects were treated within the generalized gradient approximation (GGA) in the Perdew–Burke–Ernzerhof (PBE) parametrization [9].

The plane-wave basis set was truncated at an energy cutoff of $E_{\text{cut}} = 650$ eV. Brillouin-zone integrations were performed using a Γ-centered Monkhorst–Pack $k$-point mesh of $12\times12\times12$. Structural relaxations were carried out until the residual forces on each atom were smaller than $10^{-2}$ eV/Å. Electronic self-consistency was achieved with an energy convergence criterion of $10^{-6}$ eV.

### 1.1 Wannier basis construction

To construct the localized basis used for the effective low-energy model, maximally localized Wannier functions (MLWFs) were generated using the Wannier90 code [8]. The

Wannier functions were constructed from a set of Bloch states obtained from VASP through the VASP2Wannier interface. Initial projections were chosen according to the relevant atomic orbitals of the correlated subspace (e.g., Lu $d$ orbitals and hydrogen $s$ orbitals). The disentanglement procedure was performed within an energy window extending from $-10$ eV to 10 eV relative to the Fermi level. The resulting Wannier functions reproduce the DFT band structure in the selected energy range with high fidelity.

## 1.2 Constrained Random Phase Approximation

The effective screened Coulomb interaction parameters were calculated within the constrained random phase approximation (cRPA) as implemented in VASP. In this framework, the fully screened interaction is obtained from the bare Coulomb interaction $v$ and the polarization function $P$ within the random phase approximation (RPA):

$$W(\omega) = \frac{v}{1 - vP(\omega)}. \tag{1}$$

In order to construct a low-energy Hamiltonian for the chosen correlated subspace, screening processes arising from electronic transitions within that subspace must be excluded. The total polarization is therefore decomposed as

$$P(\omega) = P_d(\omega) + P_r(\omega), \tag{2}$$

where $P_d$ corresponds to transitions within the correlated subspace and $P_r$ represents the remaining screening channels. The partially screened interaction defining the effective Hubbard interaction is then

$$W_r(\omega) = \frac{v}{1 - vP_r(\omega)}. \tag{3}$$

The static limit of this interaction ($\omega \to 0$) provides the effective Hubbard interaction parameters used in model Hamiltonians:

$$U_{ijkl} = \langle w_i w_j | W_r(\omega = 0) | w_k w_l \rangle, \tag{4}$$

where $w_i$ are the Wannier orbitals spanning the correlated subspace.

The matrix elements of the screened interaction were evaluated in the Wannier basis constructed from the DFT Bloch states. The calculated interaction matrix was subsequently used to extract the intra-orbital Hubbard interaction $U$, the inter-orbital interaction $U'$, and the Hund's coupling $J$.

## 1.3 Density Functional Theory + Dynamical Mean Field Theory methodology

To capture the effects of strong electronic correlations beyond conventional density functional theory (DFT), we employed dynamical mean-field theory (DMFT) in combination

with a Wannier-based low-energy Hamiltonian. The DFT+DMFT calculations were performed using the DMFTwDFT framework [11], which interfaces DFT codes with many-body solvers via Wannier90.

In this approach, a correlated subspace is constructed from a set of localized Wannier orbitals, which provide an effective low-energy Hamiltonian,

$$H = H_{\mathrm{DFT}} + H_{\mathrm{int}} - H_{\mathrm{DC}}, \tag{5}$$

where $H_{\mathrm{DFT}}$ is the Wannier-projected non-interacting Hamiltonian, $H_{\mathrm{int}}$ describes the local Coulomb interactions within the correlated subspace, and $H_{\mathrm{DC}}$ is the double-counting correction accounting for interactions already included at the DFT level.

Within DMFT, the lattice problem is mapped onto a self-consistent quantum impurity problem, in which the local Green's function is given by

$$G_{\mathrm{loc}}(i\omega_n) = \sum_{\mathbf{k}} \left[(i\omega_n + \mu)\mathbb{I} - H_{\mathrm{DFT}}(\mathbf{k}) - \Sigma(i\omega_n)\right]^{-1}, \tag{6}$$

where $\Sigma(i\omega_n)$ is the local self-energy and $\mu$ is the chemical potential. The impurity problem is solved iteratively to obtain $\Sigma(i\omega_n)$, which is then used to update the lattice Green's function until convergence is achieved.

The interacting Hamiltonian $H_{\mathrm{int}}$ is expressed in terms of effective Hubbard interactions ($U$, $U'$, $J$), obtained from constrained random phase approximation (cRPA) calculations. The DMFT impurity problem was solved using the continuous-time quantum Monte Carlo (CTQMC) method in the hybridization expansion, as implemented in the DMFTwDFT package.

Charge self-consistency between DFT and DMFT was achieved by computing the DMFT density matrix and feeding it back into the DFT calculation, ensuring a consistent treatment of electronic correlations at both the one-particle and many-body levels [11]. This iterative loop continues until convergence in both the charge density and self-energy is reached.

Finally, the momentum-resolved spectral function was obtained from the analytically continued Green's function,

$$A(\mathbf{k},\omega) = -\frac{1}{\pi}\mathrm{Im}\,\mathrm{Tr}\left[(\omega + \mu)\mathbb{I} - H_{\mathrm{DFT}}(\mathbf{k}) - \Sigma(\omega)\right]^{-1}. \tag{7}$$

## 1.4 Correlation diagnostics: self-energy, quasiparticle weight, and local moments

We characterize the strength of electronic correlation using three complementary quantities extracted from the converged DMFT solution: the quasiparticle mass enhancement obtained from the Matsubara self-energy, the per-atom occupation, and the instantaneous local moment obtained from the CTQMC impurity solver's atomic-state probabilities.

**Quasiparticle weight and mass enhancement.** Near the Fermi level, the real part of the self-energy can be linearized as $\mathrm{Re}\,\Sigma(\omega) \approx \mathrm{Re}\,\Sigma(0) + \omega\,\partial_\omega \mathrm{Re}\,\Sigma(\omega)|_{\omega=0}$, which defines

the quasiparticle weight

$$Z = \left(1 - \frac{\partial \operatorname{Re}\Sigma(\omega)}{\partial\omega}\bigg|_{\omega=0}\right)^{-1}. \tag{8}$$

For local/momentum-independent self-energy, $Z$ is exactly the inverse mass enhancement, $m^*/m = 1/Z$. We have:

$$\frac{m^*}{m} = 1 - \frac{\partial \operatorname{Im}\Sigma(i\omega_n)}{\partial\omega_n}\bigg|_{\omega_n\to 0}, \tag{9}$$

**Occupation and local moment.** The per-atom occupation $n$ of each correlated orbital is obtained from the impurity solver and cross-checked against the trace of the local Green's function. For a single-orbital impurity we have the probabilities of the empty, singly-occupied (spin-up and spin-down), and doubly-occupied configurations, $P_{\text{empty}}$, $P_\uparrow$, $P_\downarrow$, and $P_{\text{double}}$, from which the occupation and the instantaneous local moment follow directly,

$$n = P_\uparrow + P_\downarrow + 2P_{\text{double}}, \qquad \langle m^2\rangle = P_\uparrow + P_\downarrow. \tag{10}$$

**Double counting.** The double-counting correction $H_{\text{DC}}$ removes the portion of the local Coulomb interaction already present, in a mean-field sense, in the DFT exchange-correlation functional, and must be subtracted to avoid counting the same interaction twice. We employ the fully localized limit (around-mean-field) form,

$$H_{\text{DC}} = U\left(n_0 - \tfrac{1}{2}\right) - \tfrac{1}{2}J\left(n_0 - 1\right), \tag{11}$$

evaluated self-consistently using the DMFT-derived occupation $n_0$ of the correlated orbital at each charge-self-consistency iteration, so that the double-counting potential tracks the converged correlated occupation rather than a fixed nominal value.

# 2 Supplementary Results

## 2.1 DFT structures and lattice trends

Figure 1 shows the Lu–H structures considered in this work. The stoichiometric end members $LuH_3$ and $LuH_2$ adopt the $Fm\bar{3}m$ structure. Hydrogen-deficient structures were generated by removing H from octahedral sites for $LuH_{3-x}$ and from tetrahedral sites for $LuH_{2-x}$, and all structures were relaxed within DFT; the optimized lattice parameters are summarized in Table S1 3.

All compositions relax to cubic or near-cubic cells, with $a = b = c$ and $\alpha = \beta = \gamma$ to within 0.02°. The lattice constant varies by less than 0.4% across the entire series and does so non-monotonically, with a shallow minimum near $LuH_{2.875}$. This near-invariance of the cell geometry is important for interpreting the interaction parameters that follow: because the lattice constant changes so little—and non-monotonically—across the series, any systematic, monotonic trend in the screened interactions with hydrogen content (Sec. 2.2.2) cannot be attributed to cell geometry and must instead be electronic in origin. We return to this point in the cRPA analysis.

## 2.2 Interaction Parameters

### 2.2.1 Dependence of interaction parameters on orbital character

We performed constrained random-phase approximation (cRPA) calculations of the effective Coulomb interaction parameters for three correlated subspaces: (i) RE-$d$ only, (ii) H-$s$ only, and (iii) the combined RE-$d$ + H-$s$ manifold. From the interaction matrices we extracted the intra-orbital Hubbard $U$, inter-orbital $u$, and Hund's coupling $J$ (Tables 5,13 and following).

Across all stoichiometries the screened intra-orbital interaction on the H-$s$ orbitals is substantially larger than on the RE-$d$ orbitals. This is the opposite of the naive expectation that the more spatially compact orbital, or the rare-earth $d$ manifold often presumed more strongly correlated, should carry the larger $U$. We find that two distinct factors must be separated to explain the hierarchy: orbital localization and screening.

The bare local Coulomb matrix element scales inversely with orbital extent, $U \sim \iint d\mathbf{r}\, d\mathbf{r}'\, |w(\mathbf{r})|^2 |w(\mathbf{r}')|^2$ $\mathbf{r}'|$, so a more localized Wannier function yields a larger bare interaction. The H-$s$ Wannier functions are indeed more compact (spread $\sim 0.63$ Å$^2$) than the RE-$d$ functions ($\sim 1.40$–$1.55$ Å$^2$), which is consistent with $U_H > U_d$ at the level of the bare interaction. However, localization alone does not control the *screened* interaction. The decisive test is the change of correlated subspace: enlarging the subspace from H-$s$ only to the combined RE-$d$ + H-$s$ manifold *increases* the H-$s$ Wannier spreads (from $\sim 0.63$ to $\sim 0.72$ Å$^2$), which on the geometric argument alone would *reduce* $U_H$. The cRPA result is the opposite: $U_H$ *increases* (from $\sim 6.1$–$6.3$ to $\sim 6.9$–$7.6$ eV). This demonstrates that the screened H-$s$ interaction is governed primarily by the available screening channels rather than by orbital extent. In cRPA the screened interaction is $W_r = [1 - vP_r]^{-1} v$, where the polarization $P_r$ excludes processes within the correlated subspace; enlarging the subspace to include RE-$d$ transfers the RE-$d$ polarization out of $P_r$, reducing the screening of $U_H$ and raising it. Because the H-$s$ spread *increases* while $U_H$ also increases, the effect cannot be geometric and must be the loss of RE-$d$ screening.

The RE-$d$ sector behaves oppositely, and this asymmetry is itself informative. Enlarging the subspace removes the H-$s$ screening of RE-$d$, yet $U_d$ barely changes (and slightly *decreases*, from 3.64–3.90 to $\approx 3.63$ eV). The H-$s$ manifold—filled, deep below $E_F$, and weakly polarizable—is a poor screener of RE-$d$, so removing it has little effect; the small residual decrease tracks a slight increase in the RE-$d$ Wannier spreads upon rehybridization. Thus RE-$d$ screens H-$s$ strongly, but H-$s$ screens RE-$d$ weakly: the screening is markedly asymmetric, consistent with RE-$d$ being the itinerant, polarizable manifold near $E_F$ and H-$s$ the localized, filled one well below it. Notably, in the combined subspace the RE-$d$ onsite $U$ becomes nearly uniform ($\approx 3.63$ eV) even though the Wannier spreads retain their 2+3 ($e_g$+$t_{2g}$) structure—a further indication that $U$ is set by screening rather than orbital extent.

Taken together, the H-$s$ states are compact and weakly screened, giving large local interactions, while the RE-$d$ states are more extended and more effectively screened despite their multi-orbital character. The hierarchy $U_H > U_d$ is therefore driven primarily by differences in screening rather than by localization alone.

Table 5 compares the interaction parameters for the Lu and La hydrides. In the dihydrides, $LuH_2$ shows larger $U$, $u$, and $J$ than $LaH_2$ in both the RE-$d$ and H-$s$ sectors. Both

Table 1: Evolution across the $LuH_x$ series: lattice constant $a$, averaged cRPA H-$s$ and RE-$d$ Hubbard $U$, and DMFT per-atom H-$s$ occupation. Octahedral occupation begins above $LuH_2$. The lattice constant is nearly invariant while $U_{\text{H-}s}$ rises by $\sim 40\%$.

| System | $a$ (Å) | $U_{\text{H-}s}$ (eV) | $U_{\text{RE-}d}$ (eV) | $n_{\text{H-}s}$ |
|---|---|---|---|---|
| $LuH_{1.75}$ | 5.027 | 6.25 | 3.74 | 1.91 |
| $LuH_2$ | 5.011 | 6.52 | 4.30 | 1.91 |
| $LuH_{2.875}$ | 5.008 | 7.76 | 4.65 | 1.88 |
| $LuH_3$ | 5.014 | 8.89 | 4.45 | 1.75 |

crystallize in $Fm\bar{3}m$ but with very different lattice constants (5.011 Å for $LuH_2$, 5.655 Å for $LaH_2$); the smaller $LuH_2$ cell yields more compact Wannier functions and correspondingly larger interactions, so the dihydride comparison is governed by the lattice constant. This geometric ordering does not, however, carry over cleanly to the trihydrides: while $U_H$ remains slightly larger for $LuH_3$ than $LaH_3$ (8.89 vs 8.40 eV), the RE-$d$ interaction *reverses*, with $LaH_3$ now larger (4.69 vs 4.45 eV) despite its much larger cell. The RE-$d$ channel in the trihydride is therefore no longer controlled by geometry alone, reflecting the change in $d$-band filling and screening octahedral occupation (Sec. S2.2).

### 2.2.2 Evolution of the screened interaction with stoichiometry

Table 1 collects the cRPA H-$s$ interaction, the DMFT hydrogen occupation, and the lattice constant across the $LuH_x$ series. The screened H-$s$ Hubbard $U$ rises substantially with hydrogen content, from 6.25 eV in $LuH_{1.75}$ to 8.89 eV in $LuH_3$—an increase of roughly 42%. Over the same range the lattice constant is nearly invariant, varying by less than 0.4% (5.027 to 5.014 Å) and non-monotonically. The enhancement of the hydrogen interaction is therefore not a geometric (lattice-volume) effect: at essentially fixed cell size, the screened interaction nonetheless grows by several eV.

The rise sets in specifically as octahedral sites become occupied. In $LuH_{1.75}$ and $LuH_2$, where only tetrahedral sites are filled, the H-$s$ interaction is comparatively small (6.25 and 6.52 eV) and the matrix is uniform across sites. Once octahedral hydrogen appears ($LuH_{2.875}$, $LuH_3$), the average $U$ jumps to 7.76 and 8.89 eV, and the H-$s$ interactions split by site type (Sec. 2.2.3). The RE-$d$ interaction, by contrast, varies only weakly and non-monotonically across the series ($U_d$ = 3.74–4.65 eV) and shows no comparable systematic trend, indicating that the strong stoichiometry dependence is specific to the hydrogen sublattice.

The screened H-$s$ Hubbard $U$ rises monotonically and substantially with hydrogen content, from 6.25 eV in $LuH_{1.75}$ to 8.89 eV in $LuH_3$ (Table 1), an increase of $\sim 40\%$. The lattice constant, by contrast, varies by less than 0.4% across these compositions and does so *non-monotonically*, with a minimum near $LuH_{2.875}$ and the trihydride larger than $LuH_{2.875}$ (Table 3). The interaction and the lattice constant thus differ in both magnitude and trend: $U_{\text{H-}s}$ grows steadily while $a$ wanders within a half-percent window. The enhancement of the hydrogen interaction is therefore decoupled from the cell geometry and is electronic in origin.

### 2.2.3 Site-resolved hydrogen interactions: tetrahedral versus octahedral

In the trihydride, where both interstitial sublattices are occupied, the screened H-$s$ interaction resolves into two site-distinct values. Identifying the sites from the Wannier-function centers, the four octahedral H-$s$ orbitals carry $U_{\rm oct} \approx 8.25$ eV and the eight tetrahedral $U_{\rm tet} \approx 9.21$ eV in $LuH_3$, a difference of $\Delta U \approx 1.0$ eV ($\sim 11\%$). The same ordering holds in $LaH_3$ ($U_{\rm oct} \approx 8.0$, $U_{\rm tet} \approx 8.8$ eV) and $LuH_{2.875}$ ($\Delta U \approx 0.7$ eV), and is absent in $LuH_2$ and $LuH_{1.75}$, where only tetrahedral sites are occupied and the H-$s$ interactions are uniform.

This difference is not a consequence of orbital extent. The tetrahedral and octahedral H-$s$ Wannier functions have nearly equal spreads, $\Omega_{\rm tet} \approx 0.51$ Å$^2$ and $\Omega_{\rm oct} \approx 0.54$ Å$^2$, differing by only $\sim 5\%$ (Table 2). For a localized $s$-like Wannier function the bare onsite Coulomb integral scales inversely with the orbital radius, $U_{\rm bare} \sim \langle r^2 \rangle^{-1/2} \sim \Omega^{-1/2}$, so the spread difference predicts a bare-interaction ratio of only

$$\frac{U^{\rm tet}_{\rm bare}}{U^{\rm oct}_{\rm bare}} \sim \left(\frac{\Omega_{\rm oct}}{\Omega_{\rm tet}}\right)^{1/2} = \left(\frac{0.54}{0.51}\right)^{1/2} \approx 1.02, \tag{12}$$

i.e. a $\sim 2\%$ geometric effect. The observed ratio of the *screened* interactions is

$$\frac{U_{\rm tet}}{U_{\rm oct}} = \frac{9.21}{8.25} \approx 1.12, \tag{13}$$

a $\sim 12\%$ effect. The geometry therefore accounts for only about one-sixth of the observed difference; the remainder must arise from screening. The two hydrogen sublattices carry essentially the same bare interaction but are screened to differing degrees, the octahedral hydrogen being the more strongly screened.

Table 2: Site-resolved H-$s$ Wannier spreads and screened onsite $U$ in $LuH_3$, in the H-$s$-only and combined RE-$d$+H-$s$ subspaces. The spreads are nearly equal between sublattices, yet $U$ differs by $\sim 1$ eV; the tet/oct ordering inverts on enlarging the subspace.

| | H-$s$ only | | combined | |
|---|---|---|---|---|
| Site | $\Omega$ (Å$^2$) | $U$ (eV) | $\Omega$ (Å$^2$) | $U$ (eV) |
| Octahedral | 0.537 | 8.25 | 0.542 | 12.5 |
| Tetrahedral | 0.513 | 9.21 | 0.535 | 12.20 |

The screening origin is confirmed directly by enlarging the correlated subspace from H-$s$ only to combined RE-$d$+H-$s$, which removes the RE-$d$ states from the cRPA screening. Both H-$s$ interactions rise by $\sim$ 3–4 eV, but the octahedral interaction rises more steeply ($8.25 \to 12.5$ eV) than the tetrahedral ($9.21 \to 12.20$ eV), *inverting* the ordering so that the octahedral interaction becomes the larger (Table 2). Throughout this change the Wannier spreads remain nearly constant ($\Omega_{\rm oct} : 0.537 \to 0.542$, $\Omega_{\rm tet} : 0.513 \to 0.535$ Å$^2$), so the several-eV shifts in $U$ cannot be geometric. Because the octahedral interaction gains more when the RE-$d$ screening channels are removed, the octahedral hydrogen is the sublattice more strongly screened by the rare-earth conduction electrons.

# 3 Hydrogen filling

To probe the origin of the hydrogen filling directly, we examine the on-site (R= 0, diagonal) elements of the Wannier Hamiltonian, $\epsilon_i = \langle w_i(\mathbf{0})|\hat{H}|w_i(\mathbf{0})\rangle$, which give the bare one-body energy of each correlated orbital before hybridization. Referencing these to the Fermi level isolates where the hydrogen states sit relative to the conduction manifold, providing a more direct measure of the hydrogen charge state than the occupation alone.

The maximally-localized Wannier functions for $LuH_3$ are well converged, with small, positive spreads (0.69 Å$^2$ for octahedral and 0.86 Å$^2$ for tetrahedral H-$s$, and $\sim 1.0$ Å$^2$ for Lu-$d$), and centers located precisely on the crystallographic sites. The H-$s$ Wannier functions resolve into two symmetry-distinct groups: eight tetrahedral and four octahedral, as expected for the fluorite trihydride.

# 4 Correlation effect calculations

### 4.0.1 Self-energy

We plot the imaginary part of the Matsubara self-energy, $-\mathrm{Im}\,\Sigma(i\omega_n)$ [Fig. S1(b)] and the analytically continued real-axis self-energy $-\mathrm{Im}\,\Sigma(\omega)$ [Fig. S1(a)]. For every orbital and every stoichiometry, $-\mathrm{Im}\,\Sigma(i\omega_n)$ extrapolates toward zero as $\omega_n \to 0$ and rises approximately linearly at low frequency. Also, $-\mathrm{Im}\,\Sigma(i\omega_n)$ shows no upturn or divergence. This is a Fermi-liquid behavior of coherent, finite-lifetime quasiparticles. This shows that LuHx are correlated metals rather than Mott insulators at the temperature considered here.

The low-frequency slope, which sets the mass enhancement $m^*/m = Z^{-1} \simeq 1 - \mathrm{Im}\,\Sigma(i\omega_0)/\omega_0$, orders the Lu-$d$ correlation strength as LuH1.75 > LuH2.875 > LuH3, i.e. the Lu-$d$ correlations *decrease* with increasing hydrogen content. The overall magnitude of $-\mathrm{Im}\,\Sigma$ is moderate (peaking at $\lesssim$ 0.25 eV), consistent with moderately—rather than strongly—renormalized quasiparticles. On the real axis, the Lu-$d$ incoherent scattering peaks lie just below $E_F$; they are broadest, with a weak above-$E_F$ shoulder, for LuH1.75 and collapse to a single sharp sub-$E_F$ peak for LuH2.875, mirroring the Matsubara-axis ordering.

The H-$s$ self-energy is roughly an order of magnitude smaller than that of Lu-$d$ for all stoichiometries, confirming weak H-$s$ correlation despite the large H-$s$ Hubbard $U$ (discussed below). Its frequency dependence is noisier at low $\omega_n$, reflecting the small signal, so the H-$s$ extrapolation is the least certain. Notably, the H-$s$ scattering *grows* with hydrogen content (LuH3 > LuH2.875 > LuH1.75) — the opposite trend to Lu-$d$ — and on the real axis its scattering is a narrow peak located essentially at $E_F$.

## 4.1 DFT and DMFT Densities of States

We compare the orbital-projected density of states (DOS) obtained from DFT (top row) and DMFT (bottom row) for the three stoichiometries in Figure S2. Within DFT, all three compositions are metallic, with substantial Lu-$d$ (blue) spectral weight at and around the Fermi level $E_F$ and H-$s$ (orange) bonding states found several eV below

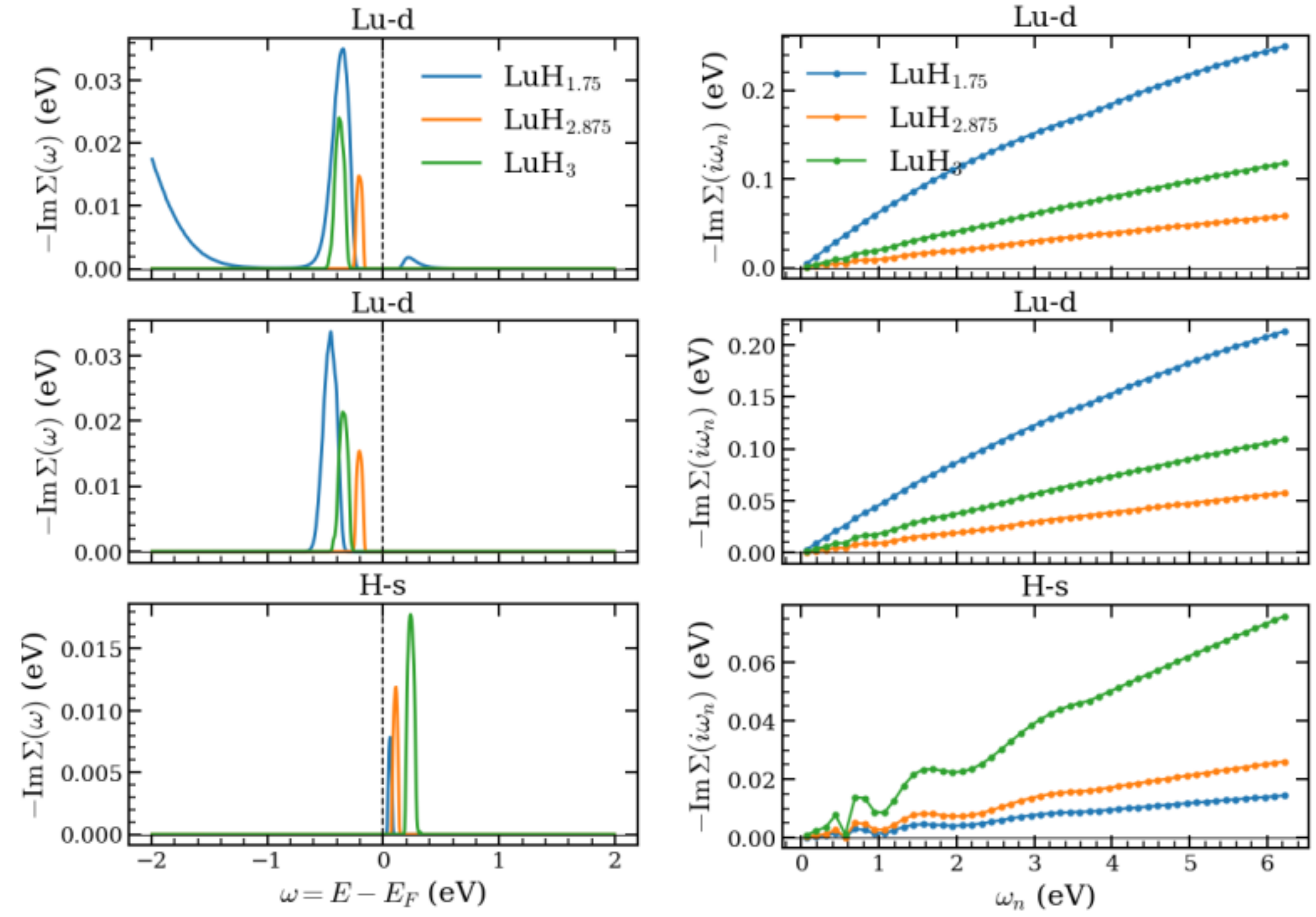


Figure S1: Imaginary part of the self-energy as calculated for various stoichiometries. Left side is analytically continued and right side is averaged over last few converged DMFT iterations

$E_F$. This DFT result is expected from mean-field itinerant-band description, with the Lu-$d$ manifold crossing $E_F$. Now, as we include dynamical correlation through DMFT, the spectral weight is redistributed in all three cases. Occupied Lu-$d$ weight is pushed to lower energy and becomes incoherent. Next, the DOS near $E_F$ is diminished, and a broad high-energy feature emerges in the range ∼1.5–4 eV, reminiscent of an upper Hubbard band. The strength of the near-$E_F$ suppression varies across the series: $LuH_{2.875}$ and $LuH_3$ show DOS minimum at $E_F$, $LuH_{1.75}$shows a strong but metallic (bad-metal-like) suppression. The H-$s$ states are mostly situated near the occupied region and are not show much weight transfer.

In every $LuH_x$ stoichiometry the H-1$s$ shell is near-full ($n_{\text{H-}s} = 1.75$–$1.91$, i.e. 88–95% occupied) and shows weaker renormalized ($m^*/m \leq 1.07$, $Z \geq 0.99$), despite carrying the *largest* cRPA interaction (up to $U_{\text{H-}s} = 8.89$ eV). The large H-1$s$ interaction manifests instead as a sizable static shift of the real part of the self-energy, $\mathrm{Re}\,\Sigma_{\text{H-}s} \sim 6$–$8$ eV, which grows with $U_{\text{H-}s}$ across the series and displaces the H-derived spectral weight from $\sim -5$ eV (DFT) to $\sim -8$ eV (DMFT).

The substoichiometric compounds exhibit multiple sharp features within a narrow energy window of ±0.5 eV around the Fermi level, in contrast to $LuH_2$, which shows a comparatively smooth DOS in this region. The DMFT spectral functions over a wider energy range (−6 to 4 eV) are shown in Fig. 2. The sharp features identified in the DFT DOS can be traced to flat bands located above $E_F$ in the non-interacting description. A closer view of the low-energy electronic structure is provided in Fig. 4. $LuH_{1.75}$ exhibits a pro-

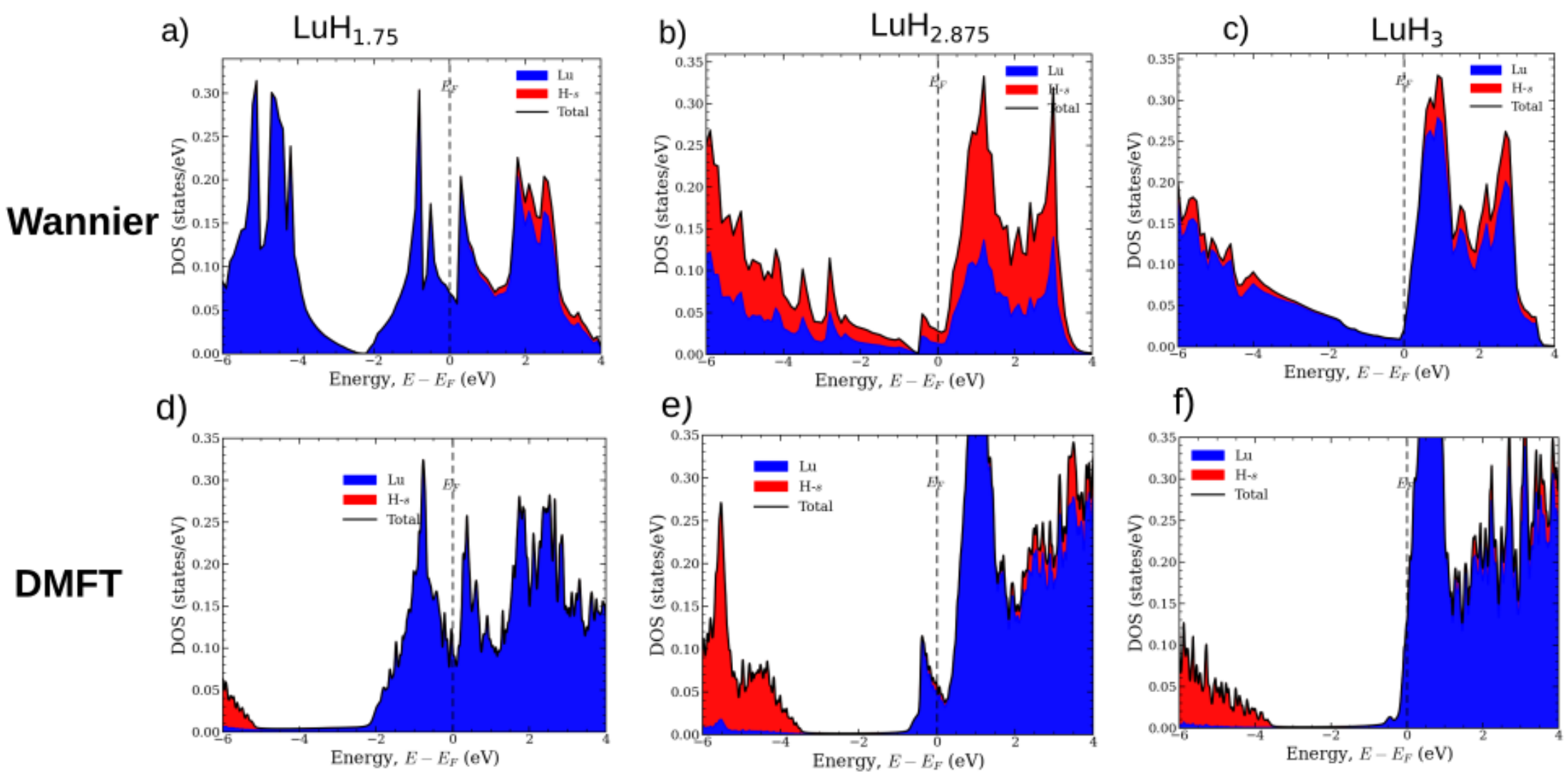


Figure S2: Comparision of DFT and DMFT densities of states of Lu-H structures calculated using cRPA calculated U values for various stoichiometries. Panels (a,d), (b,e), and (c,f) correspond to $LuH_{1.75}$, $LuH_{2.875}$, and $LuH_3$, respectively.

nounced quasiparticle peak near −0.3 eV. In addition, the DMFT spectra clearly resolve the crystal-field splitting between the $e_g$ and $t_{2g}$ manifolds. DFT DOS shows $LuH_3$ as semi metallic whereas DMFT shows $LuH_3$ as H-dominated metal.

To isolate the variable controlling the H-1$s$ correlation, we repeated the $LuH_3$ and $LaH_3$ calculations with the H-1$s$ interaction raised to $U_{\text{H-}s} = 14$ eV, well above the cRPA values, holding all else fixed. In $LuH_3$, where H-1$s$ is near-full, this 58% increase in $U$ changes the H-1$s$ quasiparticle weight ($Z = 0.99 \rightarrow 0.83$). H-1$s$ correlation is controlled by the orbital filling, not by the magnitude of $U$: on-site interaction can renormalize a half-filled H-1$s$ shell but is inactive on a near-full one, regardless of its strength. Even at $U_{\text{H-}s} = 14$ eV, however, $LuH_3$ and $LaH_3$ remains a correlated metal, not an insulator.

### 4.1.1 Quasiparticle renormalization and local moments

We quantify the local electronic correlation in each manifold using three complementary diagnostics extracted from the DFT+DMFT solution: the quasiparticle mass enhancement $m^*/m = 1/Z$, obtained from the low-frequency slope of the Matsubara self-energy $m^*/m = 1 - \partial\,\text{Im}\Sigma(i\omega_n)/\partial\omega_n|_{\omega_n\rightarrow 0}$; the per-atom occupation $n$; and the instantaneous local moment $\langle m^2\rangle$, obtained directly from the CTQMC atomic-state probabilities. The mass enhancement is a low-energy Fermi-liquid quantity and is the most sensitive to statistical noise; the occupation and local moment are static expectation values and are correspondingly robust. We therefore treat the local moment and occupation as the primary correlation diagnostics and report $m^*/m$ only where the self-energy slope is resolvable above the statistical noise (verified by a Fermi-liquid check on the low-frequency self-energy and a fit-window sensitivity analysis). Tables 15 and 16 collect the results for the H-$s$ and RE-$d$ manifolds, respectively.

## 4.2 Extended discussion of cRPA screening trends

**Screening versus localization in the cRPA literature.** That screening, rather than orbital extent, controls interaction trends is established for the transition-metal oxides. Across the $SrMO_3$ series the $t_{2g}$ interaction *decreases* with atomic number, contrary to the atomic localization trend, because increasing screening overcompensates the increasing orbital localization [14]. The decisive separation of the two effects—bare interaction (orbital extent) versus screening—was demonstrated for metal surfaces, where the variation of $U$ was shown not to arise from changes in the Wannier-function spread (the bare interaction being essentially unchanged) but from changes in screening, such that $U$ can even decrease where a simple coordination argument predicts an increase [10]. Our finding that $U_H$ rises even as the H-$s$ Wannier functions become more extended is an instance of the same screening-over-localization principle, and we use the same logic to attribute the H-$s$ interaction differences to screening.

**Dihydride Lu/La comparison.** The $LuH_2$/$LaH_2$ comparison, by contrast, is governed primarily by structure. The much smaller $LuH_2$ lattice constant yields more localized Wannier functions for both RE-$d$ ($\sim 1.40$ vs $\sim 2.17$ Å$^2$) and H-$s$ ($\sim 0.63$ vs $\sim 0.78$ Å$^2$), accounting for the larger interactions in $LuH_2$. This geometric ordering is specific to the dihydride: in the trihydride the RE-$d$ interaction is larger for $LaH_3$ despite its larger cell, indicating that octahedral occupation introduces screening and filling changes that override the geometric trend in the $d$ channel. The rare-earth species also differ in $f$-shell occupation (filled for Lu, empty for La), which may further modify the local potential and screening; our calculations do not isolate this from the lattice-constant effect, and we do not assign it a definite role.

**Relation to the experimental lattice contraction.** The interpretation of the stoichiometry trend aligns with the long-standing anionic picture of hydrogen in the rare-earth hydrides [12, 3] and with the experimental observation that heavy rare-earth $\beta$-$RH_{2+x}$ contracts with increasing hydrogen content, attributed to increasing ionic character of the octahedral hydrogen [13]. Our fixed-composition calculations are not designed to reproduce this small structural contraction directly—the lattice constants across our series vary by less than half a percent and non-monotonically—and we therefore treat the structural trend as experimental context. What our calculations establish is the *electronic* side of the same picture: at fixed geometry, the screened hydrogen interaction tracks the anionic conversion of the octahedral sublattice. The site-resolved interactions reinforce this reading: the octahedral hydrogen, which carries the additional anionic charge, is the sublattice whose interaction is most sensitive to conduction-electron screening, as shown by the inversion of the tetrahedral/octahedral $U$ ordering when the RE-$d$ screening channels are removed.

**Filling control of the $d$ manifold; scope of the analogy.** The hydrogen reservoir sets the $d$-band filling: adding octahedral hydrogen withdraws charge from the $d$ band, reducing $n_d$ from $\approx 0.83$ in the dihydrides to $\approx 0.56$ in $LuH_3$, and the $d$ correlation tracks this dilution, weakening from $m^*/m \approx 1.09$ to 1.04, with the lowest $d$ filling ($LuH_{2.875}$, $n_d \approx 0.27$) showing no resolvable enhancement. We stress that the filling-control principle at work here is not an orbital-selective Mott transition—there is only

a single hydrogen $s$ orbital per site, with no intra-shell bandwidth disparity to drive orbital-selective localization—but a demonstration that the underlying principle extends to, and is most transparently seen in, a single-orbital host.

## 4.3 Construction of energy-distribution curves

Calculated energy-distribution curves are vertical cuts of the converged spectral function $A(\mathbf{k},\omega) = -\mathrm{Im}\, G(\mathbf{k},\omega)/\pi$ at fixed momentum, evaluated on the same Wannier Hamiltonian and self-energy used for Fig. S3. Because the self-energy is momentum independent, these require no additional DMFT iterations: the Green's function is re-evaluated on the desired $\mathbf{k}$ list at fixed $\Sigma(\omega)$ and chemical potential.

Three operations convert $A(\mathbf{k},\omega)$ into a measured-like intensity, applied in the order

$$I(\mathbf{k},\omega) = \left[\tilde{A}(\mathbf{k},\omega)\, f(\omega, T_{\mathrm{exp}})\right] \otimes R(\omega), \tag{14}$$

where $f$ is the Fermi function at the measurement temperature $T_{\mathrm{exp}} = 20$ K, $R$ is a Gaussian of 30 meV full width at half maximum representing the analyzer resolution, and $\tilde{A}$ denotes $A$ after momentum broadening. For out-of-plane cuts the momentum broadening is a Lorentzian of half width 0.15 Å$^{-1}$, corresponding to the momentum uncertainty $\Delta k_z \simeq 1/\lambda$ associated with a photoelectron escape depth $\lambda \approx 5$ Å; it is applied by convolution along the cut, with the spectral function mirror-extended about each endpoint so that the kernel is not truncated at the segment edges. In-plane cuts are shown without momentum broadening. All three parameters are fixed by the experimental conditions and none is adjusted to improve agreement. Peak positions quoted in the main text are obtained by parabolic refinement of the maximum of the raw $A(\mathbf{k},\omega)$ within an energy window bracketing a single band, before the Fermi cutoff and resolution are applied, since both displace the apparent maximum.

Out-of-plane curves are computed along the pure $k_z$ line at fixed in-plane momentum $\mathbf{k}_\parallel = (\frac{1}{2}, \frac{1}{2})$, i.e. the $M$–$R$ segment of the simple-cubic zone. The dispersion of the $-0.6$ eV feature is quoted over the full segment rather than at a single point: since $E(k_z) = E(-k_z)$, the derivative $\partial E/\partial k_z$ vanishes at both the zone centre and the zone boundary, so curves selected from the immediate vicinity of either endpoint appear flat by symmetry irrespective of the dispersion in between.

## 4.4 Limitations of the comparison of EDC

The measured and calculated curves differ in three respects that bear on the comparison. Energetically, the measurement was performed at 15–30 K with $\sim 30$ meV analyzer resolution, whereas the impurity problem is solved at $T = 0.02$ eV; the calculated linewidths are therefore upper bounds, and analytic continuation of $\Sigma(i\omega_n)$ limits the precision of feature positions near $E_F$ to a few tens of meV. In momentum, the calculation is evaluated at definite $\mathbf{k}$, whereas photoemission integrates over an intrinsic $k_z$ uncertainty set by the escape depth and, in these samples, over an additional in-plane spread arising from their multi-domain character, for which high-symmetry directions could not be established. Neither the in-plane origin nor the $k_z$ of the measured curves is therefore known, and no momentum-resolved agreement is claimed. In intensity, the measured

curves include photoemission matrix elements, which are orbital-, geometry-, and photon-energy-dependent, together with an inelastic background that grows with binding energy; neither is contained in $A(\mathbf{k}, \omega)$, which is why the calculated curves return to zero at high binding energy while the measured ones do not.

A further caveat concerns the ordered arrangement of octahedral hydrogen in our supercell. The energy of the octahedral-derived states is robust, but their residual dispersion and fine structure are artifacts of that ordering: in a disordered sample the same states would appear as an incoherent resonance at similar energy rather than as coherent mini-bands. Any residual excess sharpness of the calculated features relative to experiment is expected on these grounds alone.

## 4.5 cRPA Interactions

The screened on-site interaction on H-$s$ substantially exceeds that on RE-$d$ across all compositions, and this hierarchy is a screening effect rather than a localization effect. The decisive evidence is the subspace test: enlarging the correlated subspace from H-$s$-only to the combined RE-$d$+H-$s$ manifold increases both the H-$s$ Wannier spread and the screened $U_H$—a combination the bare-interaction (geometric) picture cannot produce, and which can only arise from the removal of RE-$d$ screening channels. This is the rare-earth-hydride analogue of the well-known result for $SrVO_3$, where including the oxygen $2p$ states in the correlated subspace nearly doubles the static $U$ by excluding the O-$p \rightarrow t_{2g}$ screening channel [1]; here the RE-$d$ manifold plays the role of the oxygen $p$ states. Likewise, cRPA results also show that screening can override orbital localization in setting interaction trends [14, 10]. We have a further discussion in Sec. SX.

The screening is noticibly asymmetric. The itinerant RE-$d$ manifold, straddling $E_F$, screens the H-$s$ interaction strongly, whereas the filled, deep-lying H-$s$ manifold screens RE-$d$ only weakly: removing H-$s$ from the screening channels leaves $U_d$ essentially unchanged. In the combined manifold the RE-$d$ interaction becomes nearly uniform even though the Wannier spreads retain their crystal-field $2+3$ structure—a further sign that the interaction is set by screening rather than orbital extent.

The screened hydrogen interaction also rises strongly with stoichiometry at nearly fixed lattice constant, identifying the trend as electronic rather than geometric. The mechanism is the progressive anionic conversion of the octahedral hydrogen above $x = 2$: each octahedral hydrogen that converts toward $H^-$ withdraws an electron from the conduction band, depleting the carriers that screen the local interactions—the same carrier-density reduction observed optically as a decreasing Drude plasma frequency [15]—so that $U_{\text{H-}s}$ rises as octahedral sites fill. The effect is specific to hydrogen; the RE-$d$ interaction shows no comparable trend. The dihydride Lu/La comparison, the role of the $f$-shell difference, and the relation to the experimental lattice contraction are discussed in Sec. SX.

## 4.6 Details on Absorptivity calculation

We finally remark on the systematic plasmon-position offset between calculation and experiment ($\sim$ 0.6 eV). Several conventional sources may contribute: the equilibrium DFT lattice constant of $LuH_2$ may differ from the experimental composition $LuH_{1.83}$ by

a few percent with corresponding effect on the Drude carrier density; convergence of the intraband contribution within `epsilon.x` requires very dense **k**-mesh sampling; and the intraband broadening parameter is treated phenomenologically rather than from first principles. We do not attribute the offset to mass renormalization, since our DMFT analysis gives $Z_{5d} \approx 0.95$ for the $LuH_2$ end member (Sec. 4.1.1), too close to unity to account for a factor of $(1.7/1.1)^2 \approx 2.4$ in the squared plasma frequency. The relative ordering and qualitative shape of the configurational dependence are robust regardless of the precise plasmon position, and these are what carry the interpretive weight of the present comparison.

## 4.7 Flat Bands, van Hove Singularity, and Comparison with Photoemission

We compare the band structures of the following cubic fluorite based structures: $LuH_{1.75}$ (one $H_{tet}$ vacancy), $LuH_2$, $LuH_{2.25}$ (one $H_{oct}$ occupancy), oct-$LuH_2$ (one $H_{tet}$ vacancy and one $H_{oct}$ occupancy) to identify the ingredients for flat bands and van Hove singularity near $E_F$. Tetrahedral vacancies provide flat bands below the Fermi level whereas octahedral occupancy pulls the van Hove singularity above M point down to the Fermi level. The structure with both $H_{tet}$ vacancy and $H_{oct}$ occupancy has van Hove singularity situated right at the Fermi level and flat bands below the Fermi level Fig. 4. The octahedral-H manifold, which the on-site Wannier alignment places at $\varepsilon - E_F = -4.21$ eV in $LuH_3$, hybridizes with the Lu-5$d$ states crossing $E_F$ in the dihydride and drives a van Hove-like feature at $M$ toward the Fermi level, while tetrahedral vacancies contribute a localized, weakly dispersing defect band. Where both coexist—the minimal model for the observed $LuH_{1.98}$ optical feature—$N(E_F)$ is enhanced relative to the clean dihydride. At the Wannier level these ingredients are present but incomplete: the narrow band is not yet pinned at $E_F$ and the saddle is not yet degenerate.

Interactions complete the construction. Since the Wannier and DMFT panels of Fig. 5 share a Hamiltonian, and since the mass enhancements extracted from the Matsubara self-energies are bounded by 1.33, the near-$E_F$ flattening is too large to be quasiparticle mass enhancement. It originates instead in the static, orbital-dependent shift $\mathrm{Re}\,\Sigma(0) - V^{\mathrm{dc}}$, which relocates the hydrogen level so that its hybridization with the Lu-$d$ manifold flattens the composite band and brings the saddle into coincidence with $E_F$. This is an interaction effect—it is absent at $U = 0$—but it is a band reconstruction rather than a correlation-induced narrowing, and its magnitude is sensitive to the double-counting prescription in the hydrogen channel, the dominant systematic uncertainty here.

Carrier doping then tunes the result. The features shift rigidly in energy at 30 meV per percent doping without displacing in momentum, electron doping moving them below $E_F$ and hole doping toward and through it. This quantifies, within a correlated framework, the proposal of Ref. [6] that aligning the flat band and saddle with $E_F$ would enhance a superconducting instability. The compositional series realizes the same tuning self-consistently, since each additional octahedral hydrogen removes $\sim$ 0.7–0.8 conduction electrons per formula unit while introducing new octahedral levels.

The photoemission comparison supports a two-component reading of both data sets. Experimentally, a sharp flat band lies immediately below $E_F$ with a light band emerging from it; in the calculation, the weakly dispersing hybridized band near $E_F$ is pro-

tected against lifetime broadening by the smallness of $|\mathrm{Im}\,\Sigma|$ at low frequency, and the deeper structure near $-0.6$ eV is dispersive and resolves into individual bands. The energy-distribution curves of Fig. S3 reproduce this correspondence: both show a sharp, momentum-independent maximum just below $E_F$, and deeper structure that disperses in plane and is comparatively flat out of plane. That the calculated deep feature appears as separate peaks rather than the single measured hump follows from momentum integration along $k_z$ over the escape-depth window and from the ordered octahedral sublattice of our supercell, which produces coherent minibands where a disordered sample would show an incoherent resonance at the same energy. Our calculations contain neither nitrogen nor oxygen, both present in the measured samples of Ref. [6]; occupation of the octahedral site by hydrogen alone therefore suffices to generate the observed phenomenology, and the all-tetrahedral $LuH_2$ calculation, in which neither feature appears, identifies octahedral occupation rather than stoichiometry as the controlling variable. Consistent with this, similar photoemission results have been obtained on nitrogen-free Lu–H crystals [7], indicating that nitrogen is not required for the flat band and saddle to appear.

The comparison is restricted to energies and qualitative band shapes. The measured crystals are multi-domain with no established high-symmetry directions, the flat-band and saddle data were taken at different and undetermined $k_z$, and our momenta refer to a supercell zone; no momentum-resolved agreement is claimed. On the theory side the impurity problem is solved at $T = 0.02$ eV, so linewidths are upper bounds, and analytic continuation limits feature positions near $E_F$ to a few tens of meV. Within these limits the picture is consistent and, notably, weakly correlated: mass enhancements nowhere exceed [1.33, local moments are fully screened, and the low-frequency self-energies are Fermi-liquid-like. The flat band and saddle are consequences of octahedral-site chemistry sharpened by an interaction-induced level shift—not manifestations of strong correlation, as has been suggested. What this leaves open is whether such features, once tuned to $E_F$, matter for pairing.

# 5 Tables

Below are the tables.

Table 3: Lattice parameters of the relaxed $LuH_x$ structures. All cells are (near-)cubic with $a = b = c$ and $\alpha = \beta = \gamma$; deviations of $\alpha$ from 90° are $\leq 0.02°$. The lattice constant varies by less than 0.6% across the full composition range.

| Composition | $a$ (Å) | $\alpha$ (°) | Volume (Å$^3$) |
|---|---|---|---|
| $LuH_{1.75}$ | 5.0272 | 90.000 | 127.055 |
| $LuH_{1.78125}$ | 5.0279 | 89.987 | 127.104 |
| $LuH_{1.8125}$ | 5.0273 | 90.015 | 127.059 |
| $LuH_{1.875}$ | 5.0288 | 90.018 | 127.179 |
| $LuH_{1.96875}$ | 5.0244 | 90.000 | 126.838 |
| $LuH_2$ | 5.0112 | 90.000 | 125.844 |
| $LuH_{2.25}$ | 5.0165 | 90.000 | 126.238 |
| $LuH_{2.875}$ | 5.0077 | 90.000 | 125.576 |
| $LuH_3$ | 5.0201 | 90.000 | 126.511 |

Table 4: Lattice parameters and volume comparision between lattice parameters optimized using DFT or DFT+U. The last row is experimentally measured lattice parameter from [5]

| Composition | $U_{Lu-d}, U_{H-s}$ (eV) | a (Å) | b (Å) | c (Å) | $\alpha, \beta, \gamma$ (°) | Volume (Å$^3$) |
|---|---|---|---|---|---|---|
| $LuH_{1.75}$ | 0.000, 0.000 | 5.0272 | 5.0272 | 5.0272 | 90.000, 90.000, 90.000 | 127.055 |
| $LuH_{1.75}$ | 2.000, 6.000 | 5.1032 | 5.1032 | 5.1032 | 90.000, 90.000, 90.000 | 132.897 |
| $LuH_2$ | 0.000, 0.000 | 5.0112 | 5.0112 | 5.0112 | 90.000, 90.000, 90.000 | 125.841 |
| $LuH_2$ | 4.300, 6.522 | 5.1771 | 5.1771 | 5.1771 | 90.000, 90.000, 90.000 | 138.759 |
| $LuH_2$ [5] | - | 5.033 | 5.033 | 5.033 | 90.000, 90.000, 90.000 | 127.491 |

Table 5: cRPA parameters in Lu-d, H-s, and combined subspaces for $LuH_x$.

| | $x$ in $LuH_x$ | In Lu-d subspace | | | In H-s subspace | | | In H-s + Lu-d subspace | | |
|---|---|---|---|---|---|---|---|---|---|---|
| | | $U_{\text{Lu-d}}$ | $u_{\text{Lu-d}}$ | $J_{\text{Lu-d}}$ | $U_{\text{H-s}}$ | $u_{\text{H-s}}$ | $J_{\text{H-s}}$ | $U$ | $u$ | $J$ |
| $LuH_3$ | 4.45 | – | – | 8.89 | – | – | 9.1 | – | – | |
| $LuH_{2.875}$ | 2.875 | 4.6462 | 1.4191 | 0.0518 | 7.7576 | 1.2005 | 0.0005 | 7.7057 | 1.4204 | 0.0236 |
| $LuH_{2.25}$ | 2.25 | 3.4047 | 0.6362 | 0.0915 | 5.7296 | 0.1233 | 0.0036 | 5.7109 | 0.6628 | 0.0541 |
| $LuH_2$ | 2 | 4.3008 | 0.8210 | 0.0952 | 6.5219 | 0.3624 | 0.0014 | 5.1666 | 0.5072 | 0.0564 |
| $LuH_{1.75}$ | 1.75 | 3.7432 | 0.6625 | 0.0951 | 6.2499 | 0.1254 | 0.0013 | 4.5974 | 0.4535 | 0.0600 |

Table 6: Screened Coulomb interaction matrix $U_{ijij}^{(H-s)}$ for the H-$s$ subspace.

$$U_{ijij}^{(H-s)} = \begin{pmatrix} 6.0695 & 0.0001 & 0.0029 & 0.0001 & 0.0029 & 0.0001 & 0.0029 \\ 0.0001 & 6.2429 & 0.0007 & 0.0001 & 0.0026 & 0.0001 & 0.0026 \\ 0.0029 & 0.0007 & 6.3170 & 0.0026 & 0.0001 & 0.0026 & 0.0001 \\ 0.0001 & 0.0001 & 0.0026 & 6.2429 & 0.0007 & 0.0001 & 0.0026 \\ 0.0029 & 0.0026 & 0.0001 & 0.0007 & 6.3170 & 0.0026 & 0.0001 \\ 0.0001 & 0.0001 & 0.0026 & 0.0001 & 0.0026 & 6.2429 & 0.0007 \\ 0.0029 & 0.0026 & 0.0001 & 0.0026 & 0.0001 & 0.0007 & 6.3170 \end{pmatrix}$$

Table 7: Wannier spreads and onsite Hubbard $U$ for the H-$s$ subspace.

| H site | Spread (Å$^2$) | $U$ (eV) |
|---|---|---|
| 1 | 0.6278 | 6.0695 |
| 2 | 0.6259 | 6.2429 |
| 3 | 0.6293 | 6.3170 |
| 4 | 0.6259 | 6.2429 |
| 5 | 0.6293 | 6.3170 |
| 6 | 0.6259 | 6.2429 |
| 7 | 0.6293 | 6.3170 |

Table 8: Screened Coulomb interaction matrix $U_{ijij}^{(Lu-d)}$ for the Lu-$d$ subspace.

$$U_{ijij}^{(Lu-d)} = \begin{pmatrix} 3.8959 & 0.3619 & 0.3622 & 0.5531 & 0.5458 \\ 0.3619 & 3.6419 & 0.4603 & 0.4851 & 0.4607 \\ 0.3622 & 0.4603 & 3.6417 & 0.4849 & 0.4606 \\ 0.5531 & 0.4851 & 0.4849 & 3.8964 & 0.3010 \\ 0.5458 & 0.4607 & 0.4606 & 0.3010 & 3.6416 \end{pmatrix}$$

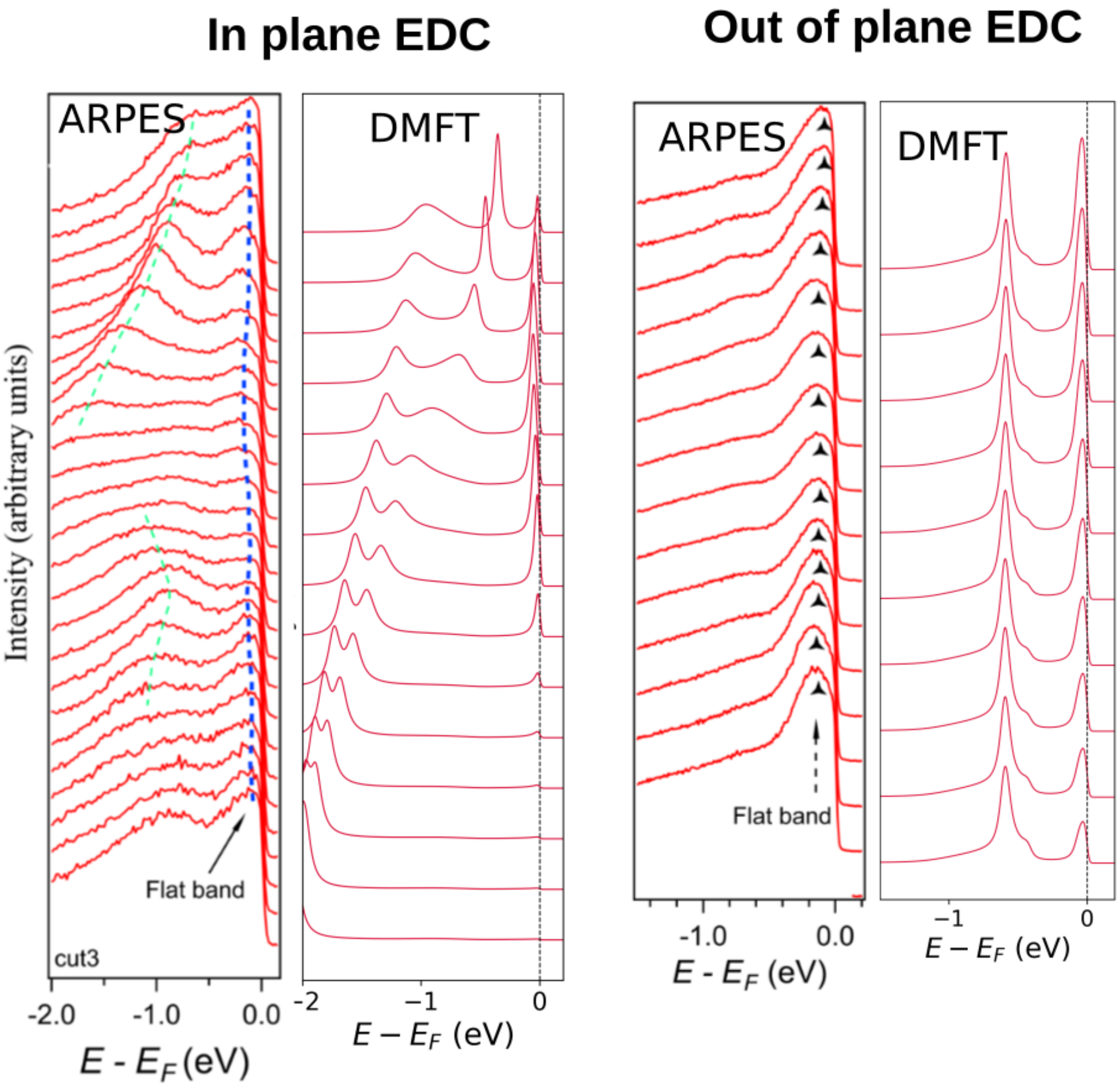


Figure S3: **Correlated electronic structure of $LuH_2$ with octahedral hydrogen: theory–experiment comparison.** In-plane and out-of-plane energy-distribution curves, adapted from Ref. [6], licensed under CC by 4.0, and calculated. Both show a sharp, momentum-independent peak near $E_F$; deeper structure is dispersive in-plane and comparatively flat out-of-plane. Momenta are not registered between calculation and experiment throughout; the comparison is made on energies and band shapes only (see text).

Table 9: Wannier spreads and onsite Hubbard $U$ for the Lu-$d$ subspace. Lu-$d$ sites show the $2+3$ orbital grouping in spreads and onsite $U$.

| Orbital | Spread (Å$^2$) | $U$ (eV) |
|---|---|---|
| 1 | 1.3801 | 3.8959 |
| 2 | 1.4409 | 3.6419 |
| 3 | 1.4408 | 3.6417 |
| 4 | 1.3800 | 3.8964 |
| 5 | 1.4410 | 3.6416 |

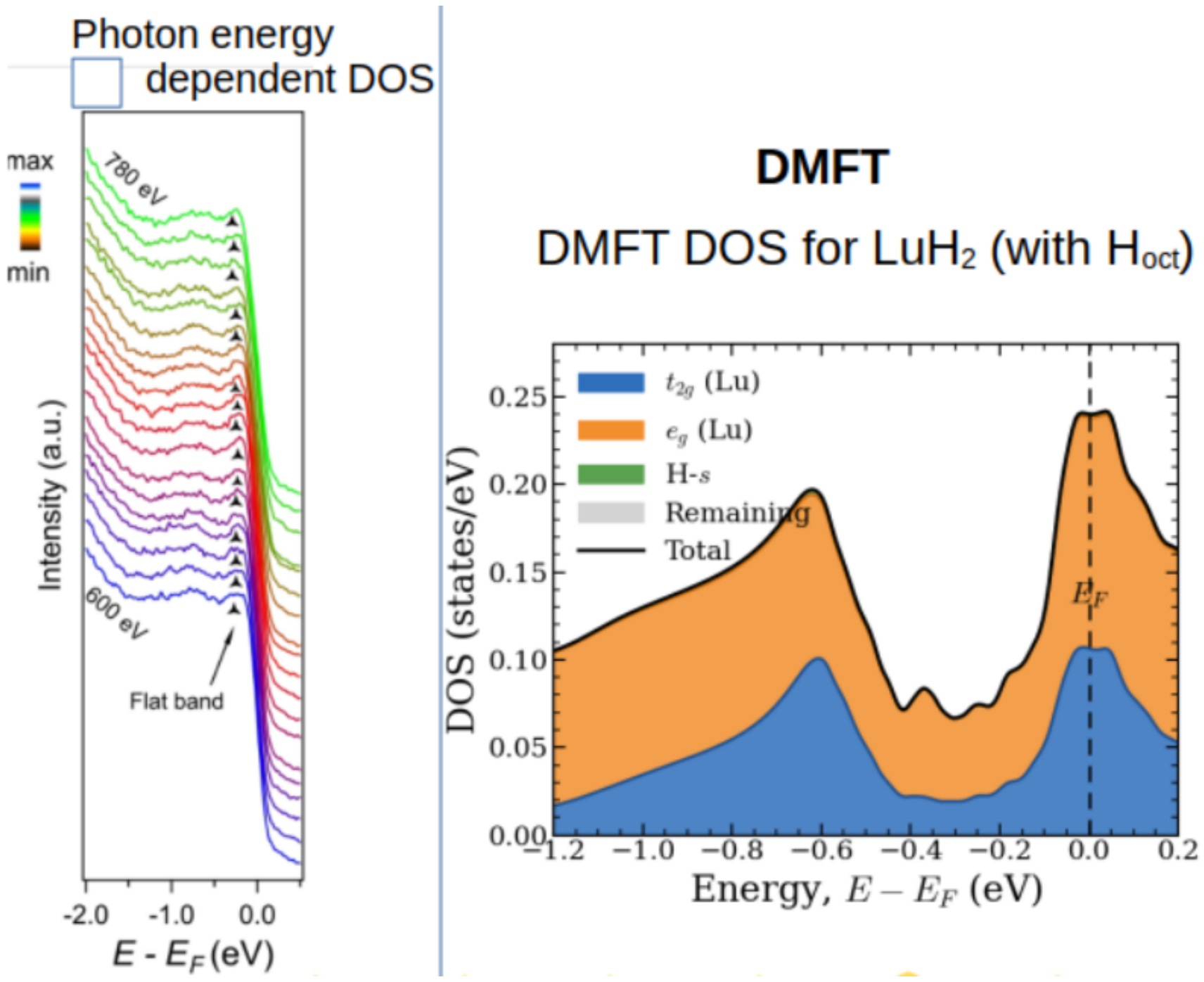


Figure S4: Photon energy dependent DOS adapted from [6] licensed under CC by 4.0 (left), and calcuated DMFT DOS (right)

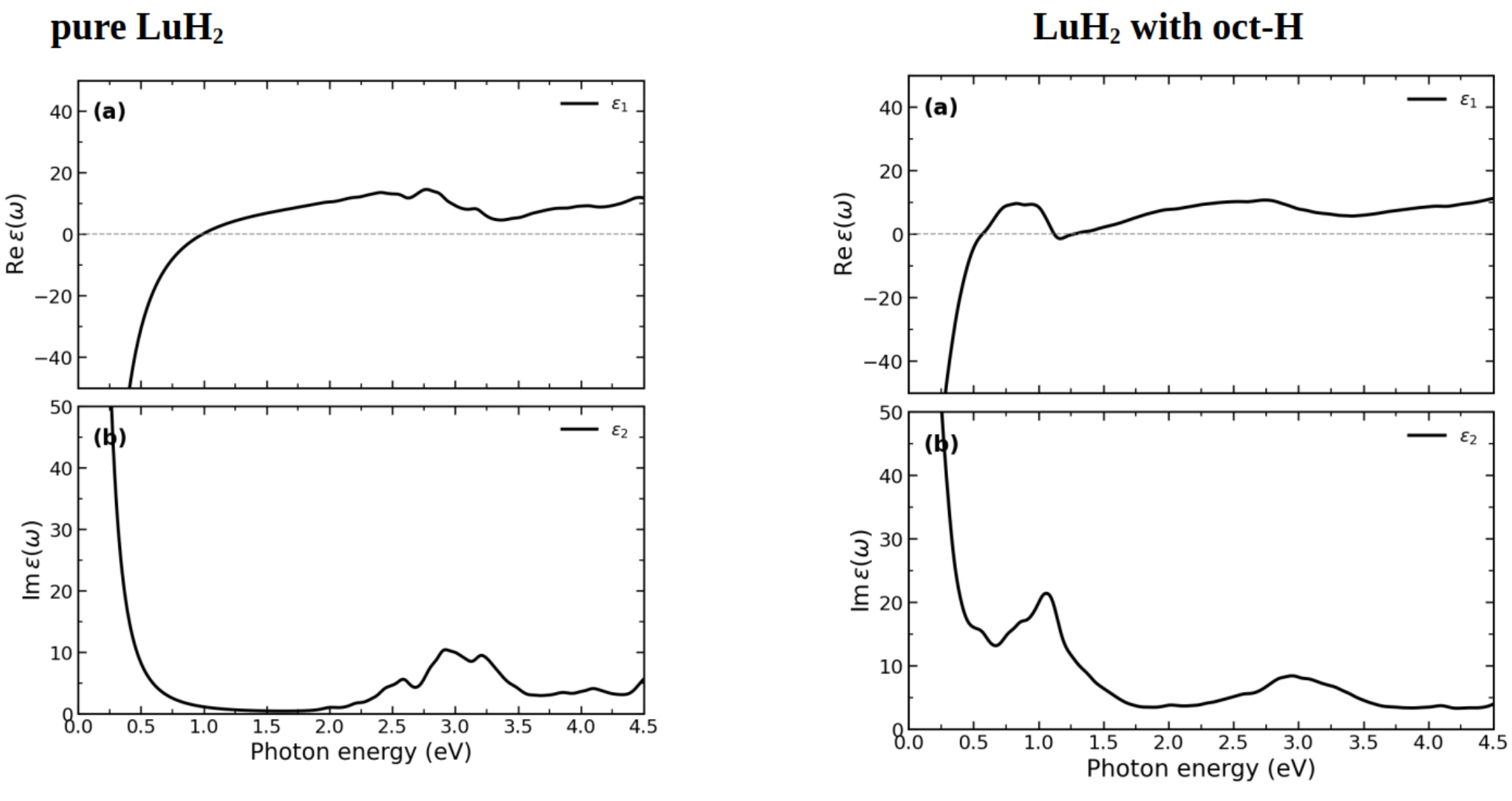


Figure S5: Calculated dielectric function of $LuH_2$ without (left) and with (right) octahedral hydrogen. (a) $\mathrm{Re}\,\varepsilon(\omega)$; (b) $\mathrm{Im}\,\varepsilon(\omega)$. In the tetrahedral-only structure, $\varepsilon_2$ falls below unity between 1.0 and 2.2 eV and $\varepsilon_1$ crosses zero within this window at 1.05 eV, giving a sharp, weakly damped plasma edge. Octahedral occupancy introduces an interband oscillator peaking at 1.05 eV inside this window, moving the zero crossing of $\varepsilon_1$ down to 0.57 eV and raising $\varepsilon_2$ at the crossing from $\approx 1$ to $\approx 15$, which damps the edge and splits the single absorptivity maximum of Fig.6 into the two observed features.

Table 10: Wannier spreads and onsite Hubbard $U$ for H-$s$ orbitals in the combined Lu-$d$ + H-$s$ correlated subspace.

| H site | Spread (Å$^2$) | $U$ (eV) |
|---|---|---|
| 21 | 0.6953 | 6.9423 |
| 22 | 0.7199 | 7.2335 |
| 23 | 0.7319 | 7.6308 |
| 24 | 0.7199 | 7.2335 |
| 25 | 0.7319 | 7.6308 |
| 26 | 0.7198 | 7.2335 |
| 27 | 0.7319 | 7.6309 |

Table 11: Wannier spreads and onsite Hubbard $U$ for Lu-$d$ orbitals in the combined Lu-$d$ + H-$s$ correlated subspace.

| Lu atom | Orbital index | Spread (Å$^2$) | Onsite $U$ (eV) |
|---|---|---|---|
| 1 | 1 | 1.42437075 | 3.6296 |
| 1 | 2 | 1.54659042 | 3.6299 |
| 1 | 3 | 1.54662202 | 3.6299 |
| 1 | 4 | 1.42443276 | 3.6301 |
| 1 | 5 | 1.54662241 | 3.6292 |
| 2 | 6 | 1.42437067 | 3.6296 |
| 2 | 7 | 1.54655558 | 3.6298 |
| 2 | 8 | 1.54660548 | 3.6298 |
| 2 | 9 | 1.42443295 | 3.6301 |
| 2 | 10 | 1.54659633 | 3.6291 |
| 3 | 11 | 1.42437190 | 3.6296 |
| 3 | 12 | 1.54655256 | 3.6297 |
| 3 | 13 | 1.54653023 | 3.6297 |
| 3 | 14 | 1.42443215 | 3.6301 |
| 3 | 15 | 1.54657519 | 3.6290 |
| 4 | 16 | 1.42437323 | 3.6296 |
| 4 | 17 | 1.54652925 | 3.6297 |
| 4 | 18 | 1.54652519 | 3.6297 |
| 4 | 19 | 1.42443113 | 3.6301 |
| 4 | 20 | 1.54657976 | 3.6290 |

Table 12: Screened interaction parameters obtained from cRPA for $LuH_x$ and $LaH_x$ for $x = 2, 3$. The table compares intra-orbital Hubbard $U$, inter-orbital interaction $u$, and Hund's coupling $J$ in the RE-$d$, H-$s$, and combined RE-$d$ + H-$s$ correlated subspaces.

| | RE-$d$ subspace | | | H-$s$ subspace | | | RE-$d$ + H-$s$ subspace | | |
|---|---|---|---|---|---|---|---|---|---|
| System | $U_d$ (eV) | $u_d$ (eV) | $J_d$ (eV) | $U_H$ (eV) | $u_H$ (eV) | $J_H$ (eV) | $U$ (eV) | $u$ (eV) | $J$ (eV) |
| $LuH_2$ | 4.3008 | 0.8210 | 0.0952 | 6.5219 | 0.3624 | 0.0014 | 5.1666 | 0.5072 | 0.0564 |
| $LaH_2$ | 3.3265 | 0.8140 | 0.0738 | 5.8031 | 0.2466 | 0.0009 | 4.8154 | 0.7786 | 0.0451 |
| $LuH_3$ | 4.45 | – | – | 8.89 | – | – | 9.1 | – | – |
| $LaH_3$ | 4.69 | – | – | 8.40 | – | – | 7.45 | – | – |

Table 13: Summary of cRPA screened interaction parameters across $LuH_x$ stoichiometries and $LaH_3$, for the three correlated subspaces (RE-$d$ only, H-$s$ only, and combined RE-$d$+H-$s$). $U$, $u$, and $J$ are the averaged intra-orbital Hubbard, inter-orbital, and Hund's coupling parameters. For the H-$s$ channel in the trihydrides, site-resolved values are given for tetrahedral (tet) and octahedral (oct) hydrogen where both are occupied. All values in eV.

| System | Subspace | Site | $U$ | $u$ | $J$ |
|---|---|---|---|---|---|
| $LuH_{1.75}$ | H-$s$ only | tet (all) | 6.25 | 0.13 | 0.001 |
| | RE-$d$ only | – | 3.74 | – | 0.09 |
| $LuH_2$ | H-$s$ only | tet (all) | 6.52 | 0.36 | 0.001 |
| | RE-$d$ only | – | 4.30 | 0.82 | 0.095 |
| | combined | – | 5.17 | 0.51 | 0.056 |
| $LuH_{2.875}$ | H-$s$ only | tet | 7.95 | – | – |
| | | oct | 7.25 | – | – |
| | | (avg) | 7.76 | 1.20 | 0.001 |
| | RE-$d$ only | – | 4.65 | – | – |
| $LuH_3$ | H-$s$ only | tet | 9.21 | – | – |
| | | oct | 8.25 | – | – |
| | | (avg) | 8.89 | 2.23 | 0.001 |
| | RE-$d$ only | – | 4.46 | 1.67 | 0.102 |
| | combined | tet | 12.20 | – | – |
| | | oct | 12.5 | – | – |
| | | (avg) | 9.19 | 2.76 | 0.045 |
| $LaH_3$ | H-$s$ only | tet | 8.76 | – | – |
| | | oct | 8.01 | – | – |
| | | (avg) | 8.51 | 2.11 | 0.001 |
| | RE-$d$ only | – | 4.69 | – | – |
| | combined | – | 7.45 | – | – |

Table 14: Site-resolved H-$s$ screened $U$ (eV) in $LuH_3$, showing the inversion of the tetrahedral/octahedral ordering upon removal of RE-$d$ screening (H-$s$-only → combined subspace).

| Subspace | $U_{\text{tet}}$ | $U_{\text{oct}}$ | $U_{\text{tet}} - U_{\text{oct}}$ |
|---|---|---|---|
| H-$s$ only | 9.21 | 8.25 | +0.96 |
| Lu-$d$ + H-$s$ | 12.20 | 12.5 | −0.3 |

Table 15: H-$s$ correlation diagnostics. Occupation $n$, atomic-state probabilities, instantaneous local moment $\langle m^2\rangle$, and quasiparticle mass enhancement $m^*/m$; local spin and charge susceptibilities $\chi_S$ and $\chi_D$ from the CTQMC solver at $\beta = 50$ eV$^{-1}$. Tetrahedral and octahedral H sites are treated as inequivalent impurities.

| System | Site | $n$ | $P_{\text{double}}$ | $\langle m^2\rangle$ | $m^*/m$ | $\chi_S$ | $\chi_D$ | $\chi_S/\chi_D$ |
|---|---|---|---|---|---|---|---|---|
| $LuH_{1.75}$ | tet | 1.833 | 0.838 | 0.157 | 1.199 | 0.0052 | 0.0047 | 1.11 |
| $LuH_2$ | tet | 1.807 | 0.814 | 0.179 | 1.221 | 0.0058 | 0.0052 | 1.12 |
| $LuH_2$ (7tet+1oct) | tet | 1.848 | 0.852 | 0.145 | 1.278 | 0.0046 | 0.0040 | 1.13 |
| | oct | 1.720 | 0.732 | 0.257 | 1.318 | 0.0129 | 0.0098 | 1.31 |
| $LuH_{2.25}$ | tet | 1.789 | – | – | – | 0.0065 | 0.0058 | 1.13 |
| | oct | 1.645 | – | – | – | 0.0169 | 0.0132 | 1.28 |
| $LuH_{2.875}$ | tet | 1.799 | 0.806 | 0.187 | 1.292 | 0.0063 | 0.0054 | 1.16 |
| | oct | 1.731 | 0.742 | 0.246 | 1.255 | 0.0121 | 0.0095 | 1.26 |
| $LuH_3$ | tet | 1.812 | 0.818 | 0.176 | 1.335 | 0.0057 | 0.0049 | 1.16 |
| | oct | 1.758 | 0.767 | 0.224 | 1.361 | 0.0104 | 0.0083 | 1.25 |
| $LaH_3$ | tet | 1.830 | 0.835 | 0.161 | 1.301 | 0.0061 | 0.0053 | 1.15 |
| | oct | 1.887 | 0.889 | 0.110 | 1.197 | 0.0055 | 0.0049 | 1.12 |

Table 16: RE-$d$ correlation diagnostics. Total $d$-shell occupation $n$ and per-orbital occupations $n_{t_{2g}}$, $n_{e_g}$ from the converged density matrix; instantaneous local moment $\langle m^2\rangle$ (group total) and mass enhancement $m^*/m$ resolved by crystal-field set; local spin and charge susceptibilities $\chi_S$, $\chi_D$ at $\beta = 50$ eV$^{-1}$. In the cubic field of the fluorite structure the $e_g$ set lies below $t_{2g}$ and is correspondingly the more occupied of the two at every composition.

| | | $n_{\text{orb}}$ | | $\langle m^2\rangle$ | | $m^*/m$ | | | | |
|---|---|---|---|---|---|---|---|---|---|---|
| System | $n$ | $t_{2g}$ | $e_g$ | $t_{2g}$ | $e_g$ | $t_{2g}$ | $e_g$ | $\chi_S$ | $\chi_D$ | $\chi_S/\chi_D$ |
| $LuH_{1.75}$ | 1.588 | 0.188 | 0.462 | 0.642 | 0.946 | 1.211 | 1.247 | 0.199 | 0.051 | 3.90 |
| $LuH_2$ | 1.428 | 0.123 | 0.489 | 0.427 | 1.001 | 1.278 | 1.295 | 0.217 | 0.047 | 4.63 |
| $LuH_2$ (7tet+1oct) | 1.417 | 0.204 | 0.360 | 0.665 | 0.753 | 1.283 | 1.320 | 0.211 | 0.039 | 5.42 |
| $LuH_{2.25}$ | 1.296 | 0.137 | 0.405 | – | – | – | – | 0.161 | 0.053 | 3.04 |
| $LuH_{2.875}$ | 0.830 | 0.122 | 0.186 | 0.487 | 0.343 | 1.204 | 1.199 | 0.060 | 0.031 | 1.93 |
| $LuH_3$ | 0.679 | 0.111 | 0.122 | 0.400 | 0.279 | 1.178 | 1.161 | 0.049 | 0.030 | 1.67 |
| $LaH_3$ | 0.531 | 0.070 | 0.105 | 0.355 | 0.175 | 1.213 | 1.203 | 0.075 | 0.041 | 1.84 |

Table 17: Charge sharing between the hydrogen and $R$-$d$ sublattices. For $R\mathrm{H}_x$ the closed-shell ionic reference has every hydrogen as $\mathrm{H}^-$, giving $\sum n_{\mathrm{H}}^{\mathrm{ionic}} = 2x$ and $n_d^{\mathrm{ionic}} = 3 - x$ per formula unit. The hydrogen deficit $\Delta_{\mathrm{H}} = 2x - \sum n_{\mathrm{H}}$ and the $d$ excess $\Delta_d = n_d - (3 - x)$ are obtained from Tables 15 and 16. The two track one another across the series with a correlation coefficient of 0.99; the residual $\Delta_d - \Delta_{\mathrm{H}} = 0.063 \pm 0.015$ is constant to within its spread and equals the deviation of the total correlated-subspace occupation from the nominal valence count $3 + x$, i.e. the weight lying outside the Wannier subspace. Charge missing from the hydrogen shell is thus recovered on the $d$ shell, establishing that the two sublattices are hybridized rather than ionically decoupled.

| System | $x$ | $\sum n_{\mathrm{H}}$ | $\Delta_{\mathrm{H}}$ | $n_d$ | $\Delta_d$ | $\Delta_d - \Delta_{\mathrm{H}}$ |
|---|---|---|---|---|---|---|
| $LuH_{1.75}$ | 1.750 | 3.208 | 0.292 | 1.588 | 0.338 | 0.046 |
| $LuH_2$ | 2.000 | 3.614 | 0.386 | 1.428 | 0.428 | 0.042 |
| $LuH_2$ (7tet+1oct) | 2.000 | 3.664 | 0.336 | 1.417 | 0.417 | 0.081 |
| $LuH_{2.875}$ | 2.875 | 5.113 | 0.637 | 0.830 | 0.705 | 0.068 |
| $LuH_3$ | 3.000 | 5.382 | 0.618 | 0.679 | 0.679 | 0.061 |
| $LaH_3$ | 3.000 | 5.547 | 0.453 | 0.531 | 0.531 | 0.078 |

Table 18: Response of the correlated subspace to rigid-band doping in $LuH_2$ (7 tet + 1 oct). One electron is added to ($+1e$) or removed from ($-1e$) the total electron count in the DMFT step at fixed structure, corresponding to ±5% doping. Occupations $n$ and local spin and charge susceptibilities $\chi_S$, $\chi_D$ from the CTQMC solver at $\beta = 50$ eV$^{-1}$. $\chi_S$ is a total-shell quantity and is therefore not directly comparable between the five-orbital $d$ manifold and the single-orbital hydrogen sites.

| Channel | Quantity | $-1e$ | undoped | $+1e$ |
|---|---|---|---|---|
| Lu-$d$ | $n$ | 1.158 | 1.417 | 1.670 |
| | $\chi_S$ | 0.1513 | 0.2108 | 0.2346 |
| | $\chi_D$ | 0.0375 | 0.0389 | 0.0391 |
| | $\chi_S/\chi_D$ | 4.04 | 5.42 | 6.01 |
| $\mathrm{H}_{\mathrm{tet}}$ | $n$ | 1.849 | 1.848 | 1.847 |
| | $\chi_S$ | 0.0045 | 0.0045 | 0.0046 |
| | $\chi_D$ | 0.0040 | 0.0040 | 0.0041 |
| | $\chi_S/\chi_D$ | 1.12 | 1.13 | 1.13 |
| $\mathrm{H}_{\mathrm{oct}}$ | $n$ | 1.725 | 1.720 | 1.720 |
| | $\chi_S$ | 0.0125 | 0.0129 | 0.0130 |
| | $\chi_D$ | 0.0096 | 0.0098 | 0.0099 |
| | $\chi_S/\chi_D$ | 1.30 | 1.31 | 1.31 |



\end{document}